\documentclass[12pt]{article}
\usepackage{amssymb}
\usepackage{amsmath}
\usepackage{latexsym}
\usepackage{epsfig}
\usepackage{graphicx}
\usepackage{subfigure}
\usepackage{wasysym}
\usepackage{threeparttable}
\usepackage{natbib}
\newcommand{\mbv}[1]{\mbox{\boldmath$#1$\unboldmath}}
\newcommand{\mbf}[1]{\mathbf{#1}}

\newcommand{\Appendix}
{%\appendix
\def\thesection{Appendix~\Alph{section}}
\def\thesubsection{A.\arabic{subsection}}
}

\def\log{\hbox{log}}

\def\boxit#1{\vbox{\hrule\hbox{\vrule\kern6pt
          \vbox{\kern6pt#1\kern6pt}\kern6pt\vrule}\hrule}}

\def\var{\hbox{var}}
\def\cov{\hbox{cov}}
\def\corr{\hbox{corr}}

\def\bse{\begin{eqnarray*}}
\def\ese{\end{eqnarray*}}
\def\be{\begin{eqnarray}}
\def\ee{\end{eqnarray}}
\def\bq{\begin{equation}}
\def\eq{\end{equation}}
\def\bse{\begin{eqnarray*}}
\def\ese{\end{eqnarray*}}

\begin{document}
\thispagestyle{empty} \baselineskip=28pt

\begin{center}
{\LARGE{\bf Semiparametric Bivariate Zero-Inflated Poisson Models with Application to Studies of Abundance for Multiple Species}}
\end{center}

\baselineskip=12pt

%%
%%
%%
%%%%%%%%%%%%%%%%%%%%%%%%%%%%%%%%%%%%%%%%%%%%%%%%%%%%%%%%%%%%%%%%%%%%%%%%
\vskip 2mm
\begin{center}
Ali Arab\footnote{(\baselineskip=10pt to whom correspondence should be addressed)
Department of Mathematics, Georgetown University,Washington DC 20057-1233, aa577@georgetown.edu},
Scott H. Holan\footnote{\baselineskip=10pt Department of Statistics, University of Missouri-Columbia,
146 Middlebush Hall, Columbia, MO 65211-6100},  Christopher K. Wikle$^2$,
Mark L. Wildhaber\footnote{\baselineskip=10pt U.S. Geological Survey,
Columbia Environmental Research Center, 4200 New Haven Road, Columbia, Missouri 65201}\\
\end{center}
%
%
%
%
%%%%%%%%%%%%%%%%%%%%%%%%%%%%%%%%%%%%%%%%%%%%%%%%%%%%%%%%%%%%%%%%%%%%%%%%

\begin{center}
{\Large{\bf Abstract}}
\end{center}

\baselineskip=12pt Ecological studies involving counts of
abundance, presence-absence or occupancy rates often produce
data having a substantial proportion of zeros.  Furthermore,
these types of processes are typically multivariate and only
adequately described by complex nonlinear relationships
involving externally measured covariates.  Ignoring these
aspects of the data and implementing standard approaches can
lead to models that fail to provide adequate scientific understanding of
the underlying ecological processes, possibly resulting in a loss
of inferential power.  One method of dealing with data having
excess zeros is to consider the class of univariate
zero-inflated generalized linear models. However, this class of
models fails to address the multivariate and nonlinear aspects
associated with the data usually
encountered in practice.  Therefore, we propose a
semiparametric bivariate zero-inflated Poisson model that takes
into account both of these data attributes.  The general
modeling framework is hierarchical Bayes and is suitable for a
broad range  of applications.  We demonstrate the effectiveness
of our model through a motivating example on modeling catch per
unit area for multiple species using data from the Missouri
River benthic fish study, implemented by the United States
Geological Survey.

%%%%%%%%%%%%%%%%%%%%%%%%%%%%%%%%%%%%%%%%%%%%%%%%%%%%%%%%%%%%%%%%%%%%%%%%
%
%
%

\baselineskip=12pt
\par\vfill\noindent
\par\medskip\noindent
\clearpage\pagebreak\newpage \pagenumbering{arabic}
\baselineskip=24pt
\section{Introduction}
The problem of having a large proportion of zero values is a
common characteristic of data obtained from environmental and
ecological studies involving counts of abundance,
presence-absence or occupancy rates \citep{clarke1988statistical, welsh1996modelling, martin2005zero, berry:wildhaber:young:2005}.
Ignoring or excluding values to facilitate the analysis of zero
inflated data can result in a loss of important information and,
thus, diminished explanatory power.  For example, when studying
abundance or presence-absence of a species in ecological
studies, having a large proportion of zero values might be an
indication that the species is rare or endangered, hard to
detect, or both. The problem of dealing with rare species and
species with a low probability of detection is extremely common
in ecological studies and so data having a preponderance of
zeros is often encountered.  Thus, standard distributions such
as Poisson, binomial and negative-binomial often fail to provide an
adequate fit.  On the other hand, one potentially appropriate
class of distributions for describing this type of data is the
class of {\it zero-inflated} distributions as they properly
account for a large proportion of zero values \citep{cohen:1963,  lambert1992zero, johnson1997discrete}.

Although it is conceivable that data having excess zeros may
come from any distribution, typically, in practice, the
distributions are discrete.  Therefore, several popular models
that account for data with excess zeros have emerged, including
the zero-inflated Poisson (ZIP), zero-inflated binomial (ZIB)
and the zero-inflated negative binomial (ZINB).  The ZIP model
is especially useful in analyzing count data with a large
number of zero observations.  However, in practice, the ZIB
model is sometimes used for cases where an upper bound exists
for the response whereas the ZINB model is sometimes used for
cases where the data are overdispersed.  Nevertheless, the ZIP
model has experienced wide-spread popularity over the last
decade and has been applied to numerous problems in horticulture \citep{hall2000zero}, manufacturing \citep{lambert1992zero}, and various other fields of study, including health operations \citep{wang2002zero}, meteorology \citep{wikle2003ZIP}, ecology \citep{welsh1996modelling, martin2005zero, ver2007space}, and fisheries biology \citep{minami2007modeling, arab2008zero, wildhaber2011}.

Despite the fact that the utility of multivariate ZIP models is
extensive, the relevant literature is somewhat limited.  Early research focused on extensions of the
univariate Poisson binomial \citep{skellam1952studies}, which is a
compound distribution of the binomial and Poisson.  However, more recently, \citet{li1999multivariate} formulate an $m$-dimensional ZIP distribution by linking all of the univariate distributions together through
one common distribution, as is done in the case of the
$m$-variate Poisson distribution \citep{johnson1997discrete}.
Moreover,  \citet{li1999multivariate} focus on the bivariate case by
deriving a bivariate ZIP (BivZIP) distribution as a mixture of
two univariate Poisson distributions and a point mass at (0,0) (i.e., a point mass when both count values equal zero).
In order to estimate these models, \citet{li1999multivariate} use
maximum likelihood. Extending this work, \citet{majumdar2010bivariate}
propose a  Bayesian BivZIP regression model with
estimation based on data augmentation. In contrast, \citet{schmidt2011} discuss models for multivariate counts observed at fixed
spatial locations based on a continuous mixture of independent
Poisson distributions.

We propose a semiparametric ZIP modeling
approach for bivariate count processes which extends the
existing bivariate zero-inflated modeling approaches to
utilization of nonlinear covariates in the model as well as
modeling zero-inflation probabilities through a multinomial logit
regression.   The modeling approach we propose produces a class
of bivariate semiparametric zero-inflated Poisson models cast in a
hierarchical Bayesian framework. Critically, the semiparametric
aspect of the proposed approach allows us to readily consider possible
nonlinear effects of covariates \citep{ruppert2003semiparametric, ruppert2009semiparametric} in a
bivariate zero-inflated setting. Finally, although the modeling framework introduced in this
paper is extremely natural for environmental and ecological
applications, the only example of such a bivariate
semiparametric modeling technique in the literature is \citet{Arab2007}.

The models we propose could potentially be considered
from either a classical or Bayesian perspective; however, as the level
of complexity increases, it is convenient (and often necessary)
to make use of the Bayesian paradigm. In this context, accounting for uncertainty in different levels of the model can
be effectively facilitated through using a hierarchical
modeling framework. For a comprehensive discussion on
hierarchical models for environmental and ecological data see
\citet{wikle2003hierarchicala}, \citet{royle2008hierarchical}, \citet{cressiewikle2011} and the references therein.

Our approach is extremely useful for developing models
of abundance in settings with multiple species, where univariate distributions are less suitable.  For example, ichthyology
(i.e., fisheries biology) is one area of ecology where modeling
counts of abundance of multiple species is prevalent.
Frequently, in this context, species are biologically related
and thus it is expected that relative abundance, as a function
of habitat, year, gear (i.e.,  type of net) used to catch the
fish, etc., will be correlated.  Therefore, a potentially
advantageous approach for modeling relative abundance is to
borrow strength across related species through the use of
multivariate distributions.

We demonstrate the effectiveness of our methodology through a motivating example in
ichthyology, namely modeling abundance of species in a
bivariate setting using fish catch data.  Here, using benthic fish data
collected by the United States Geological Survey (USGS) on the
Missouri River \citep{berry:wildhaber:young:2005}, we illustrate the usefulness of our framework by
modeling the catch per unit area (CPUA), while
determining which factors are related to the zero-inflation
probability for a given fish species and which factors are
related to catch rates. Critically, these goals are
accomplished while accounting for the dependence between
different species.

The remainder of this paper is organized as follows.  Section~\ref{sec:MotEx}
describes our motivating example, the Missouri River benthic
fish study. Semiparametric BivZIP
models are presented in Section~\ref{BVZIP}.  Section~\ref{application} applies the
proposed model to our motivating example, modeling CPUA for benthic fish on the Missouri River.  Finally, Section~\ref{sec:concl} contains discussion.  Derivation of all full conditional distributions and details surrounding our Markov chain Monte Carlo (MCMC) algorithm are left to the Appendix.

\section{Missouri River Benthic Fish Study}\label{sec:MotEx}
\begin{figure}
\centerline{
\includegraphics[height=6in,width=6.75in]{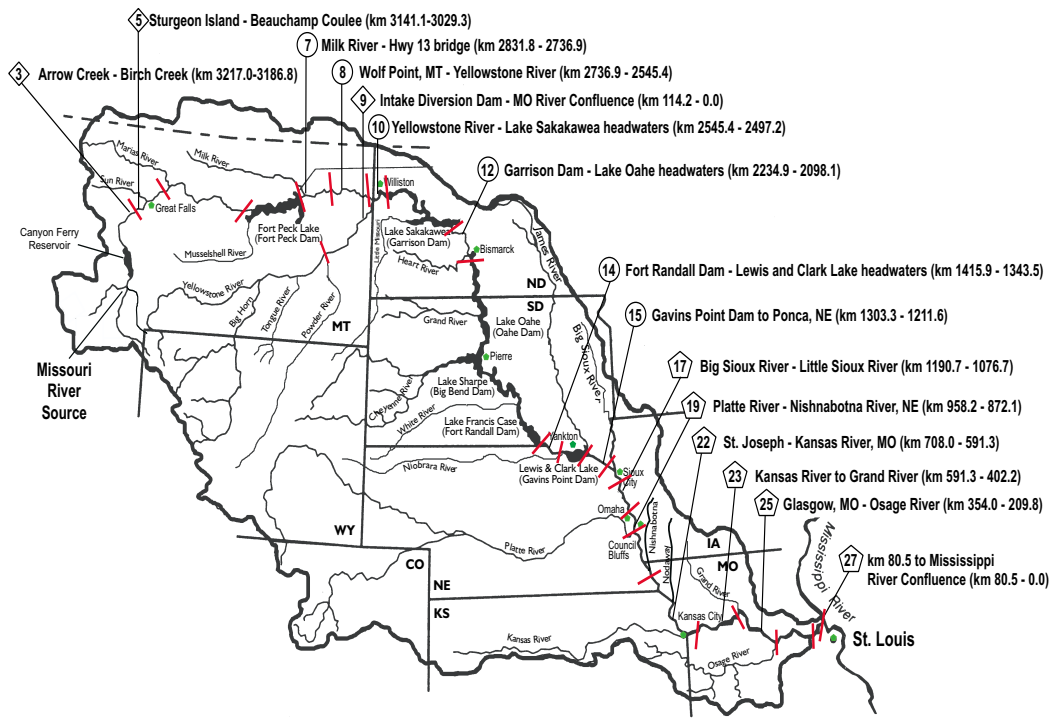}}
\caption{ \baselineskip=10pt Missouri River Benthic Fishes Study area from Montana to
the confluence of the Missouri River with the Mississippi River in Missouri, USA
(\cite{berry:wildhaber:young:2005}; $\diamondsuit=$ Least-Altered (LA), $\bigcirc=$ Inter-Reservoir
(IR), $\pentagon=$ Channelized (CH) Segments).} \label{fig:zones}
\end{figure}

In 1995, USGS and the Montana Department of Fish, Wildlife, and
Parks commenced a study to look at benthic fishes in the
warm-water portion of the Missouri River system \citep{BerryYoung2001, berry:wildhaber:young:2005}.  The Missouri River (USA) extends
3,764 kilometers from southwest Montana to the Mississippi River and
contains several species of benthic fish (Figure~\ref{fig:zones}). Benthic fish are fish
that live or feed on the bottom of the river and are of
particular interest because of their sensitivity to changes in
habitat.  The main goal of the study was to evaluate the
status, distribution and habitats associated with various
benthic fish in the Missouri River for the purpose of providing
vital information necessary for improvement in their
management.

Included in this study were 26 different benthic fish species; however, in Section \ref{application} we
will focus only on two species, common carp ({\it Cyprinus
carpio}) and channel catfish ({\it Ictalurus punctatus}) which
are two generalist species with overlapping habitat associations. In this study, fisheries biologists divided
the Missouri River into three zones and each zone was divided
into segments; the upper zone or ``least-altered zone'' (LA)
included segments 3, 5, and 9, the middle or
``inter-reservoir'' zone (IR) included segments 7, 8, 10, 12,
14 and 15, and the lower or ``channelized zone'' (CH) included
segments  17, 19, 22, 23, 25 and 27 (see Figure~\ref{fig:zones} and \citet{wildhaber2011} for
further details). Although there were twenty seven
segments included in the study design, due to financial and
administrative constraints, only fifteen segments were sampled
during each of the three years of the study considered here.

Common to fisheries field studies, the data considered here are
based on multiple gears in addition to consisting a large proportion of
zeros, thus complicating analysis.  Using standard parametric
statistical methods on data from each gear separately, \citet{berry:wildhaber:young:2005} excluded several river segments and macrohabitats
from the analysis due to a high percentage of zero observations
(i.e., violation of normality and homogeneity of variance
assumptions). As a result, the previous analysis may have
imposed limitations on the inferential scope, making
comprehensive conclusions about the complete study domain inaccessible. In particular, the analyses conducted by
\citet{berry:wildhaber:young:2005} were limited due to several factors.
First, the authors aggregated the data to larger spatial and
temporal scales in order to alleviate a large percentage of zero
observations and thus potentially caused a loss in information.
Second, separate analyses were conducted for each gear
rendering inference on gear intractable. In addition, because this
study considered each species separately, using a univariate
analysis, the ability to draw conclusions regarding correlation between various species was necessarily limited.

Our goal is to develop and
implement a modeling framework which will allow for meaningful
ecological interpretations based on the model results while
concurrently raising the predictive precision of the model in
the presence of realistic constraints.
Ultimately, similar to the zero-inflated modeling approach
discussed in \citet{wildhaber2011}, our approach can identify the type of gear, macrohabitat, segment, year and
physiochemical characteristics that explain where certain species are
most likely to populate.  The approach we propose for this data is a hierarchical Bayesian semiparametric
BivZIP model. Specifically, the model we develop incorporates
those parameters that help explain the mean fish count as well
as those that explain the zero-inflation probability (i.e.,
excess zero observations), while accommodating nonlinear
relationships and borrowing strength across \textit{``similar"} species.

The data used in our analysis was collected over the three
study years 1996 to 1998 (see \citet{berry:wildhaber:young:2005} for a
comprehensive discussion).  Channel catfish and common carp were chosen because, in part,
they reflect two commercially important species that can be
found in many habitats within different rivers, including the
Missouri River. Biologically, a study of these two species, in a
bivariate setting, is of interest due to overlapping habitat use \citep{berry:wildhaber:young:2005}. Based on criteria similar to \citet{berry:wildhaber:young:2005}, we excluded several outlier observations as well as data for segments 7, 8, 10, 12, and 14 from the analysis, due to extremely low levels of catch or none at all, for at least one of the two species.  Even though this resulted in an analysis that only used 960 of the 1477 total observations available, our analysis is still able to include more segments than the previous analysis of \citet{berry:wildhaber:young:2005}, since our modeling approach directly accommodates a significantly higher percentage of zero values than models following standard distributions\footnote{In a univariate setting, \citet{berry:wildhaber:young:2005} excluded segments 3, 5, 7, 8, 10, 12, and 14 for channel catfish.}.

Although, the resulting data contains many observations greater than zero it also contains a substantial portion of zero observations. Specifically, after removing the aforementioned segments, the observations for channel catfish and common carp included 51\% zeros and 67.7\% zeros respectively. In the bivariate setting, the percentage of zeros for both species is 39.7\%, approximately 28\% of the samples include a non-zero catch for channel catfish and zeros for common carp, approximately 11\% of the samples include zeros for channel catfish and a non-zero catch for common carp, and finally 21.25\% non-zero observations for both species.

Based on biological considerations, ecologists expect that these two species use
similar habitats and, as such, it is sensible that
abundance of each species, as a function of gear, segment,
macrohabitat and year, should be related.  In addition, exploratory data analysis, combined with current species specific knowledge, suggests that some of the
explanatory variables are nonlinearly related to the Poisson log-intensity parameters.
Consequently, there is sufficient motivation for considering
a semiparametric BivZIP modeling approach in this context.

\section{Semiparametric Bivariate Zero-Inflated Models}\label{BVZIP}
It is natural to model correlated counts using a bivariate
discrete distribution such as a bivariate Poisson distribution.  However, the computational difficulties involved
in fitting such models have traditionally deterred researchers
from using such an approach.  Recent advances in hierarchical
Bayesian modeling and, specifically, the improvement of
computational methods such as MCMC,
have provided mechanisms for easy implementation of bivariate
discrete distributions such as the bivariate Poisson (e.g., see \citet{tsionas2001bayesian}, \citet{ntzoufras2009} and \citet{majumdar2010bivariate}).  Although the bivariate Poisson distribution can be formulated from several directions \citep{kocherlakota1992bivariate, schmidt2011}, the formulation chosen here is a natural extension to the univariate Poisson distribution
that allows for correlation among the response variable for the
two populations under consideration \citep{li1999multivariate}.

Similar to \citet{li1999multivariate}, we let $Y_{1j}$ and $Y_{2j}$ denote the
$j$-{th} observation from the first and second population, respectively.  Then, for
$j=1,\ldots,n$,
\begin{eqnarray*}
Y_{1j}&=& Z_{1j}+Z_{3j},\\
Y_{2j}&=& Z_{2j}+Z_{3j},
\end{eqnarray*}
where $(Y_{1j},Y_{2j})^\prime\sim\hbox{BivPois}(\lambda_{1j},\,
\lambda_{2j},\,\lambda_{3j})$ and $Z_{1j}$, $Z_{2j}$ and
$Z_{3j}$ are mutually independent Poisson random variables with
intensity parameters $\lambda_{1j}$, $\lambda_{2j}$ and
$\lambda_{3j}$, respectively \citep{kocherlakota1992bivariate, johnson1997discrete}.  Assuming $Y_{1j}$ and $Y_{2j}$
are variables from a bivariate Poisson distribution, the
covariance between $Y_{1j}$ and $Y_{2j}$ is given by
\begin{eqnarray*}
 \cov(Y_{1j}, Y_{2j})=\cov(Z_{1j}+Z_{3j},Z_{2j}+Z_{3j})
 =\var(Z_{3j})= \lambda_{3j},
\end{eqnarray*}
and, thus, the correlation coefficient between $Y_{1j}$ and $Y_{2j}$
is
\begin{eqnarray*}
 \corr(Y_{1j}, Y_{2j})=\frac{\lambda_{3j}}{\sqrt{(\lambda_{1j}+\lambda_{3j})(
 \lambda_{2j}+\lambda_{3j})}}.
\end{eqnarray*}
The joint probability mass function is given by
\begin{eqnarray}
 \hbox{P}(Y_{1j}=y_{1j},
 Y_{2j}=y_{2j})=\exp\{-(\lambda_{1j}+\lambda_{2j}+\lambda_{3j})\}\sum_{i=0}^{\min(y_{1j},y_{2j})}
 \frac{\lambda_{1j}^{y_{1j}-i}\lambda_{2j}^{y_{2j}-i}\lambda_{3j}^{i}}{(y_{1j}-i)! (y_{2j}-i)!i!}. \label{eq:pmf}
\end{eqnarray}

To construct a BivZIP model,
we consider a mixture of a point mass at $(0,0)$, two
univariate Poisson random variables, and a bivariate Poisson random variable.
Then
\begin{eqnarray}
(Y_{1j}, Y_{2j})' \sim
\left\{
  \begin{array}{ll}
    (0, 0) & \;\;\text{w.p.} \;\;\;\;p_{0j}, \\
   (\hbox{Pois}(\lambda_{1j}+\lambda_{3j}),0) & \;\;\text{w.p.} \;\;\;\; p_{1j}, \\
   (0, \hbox{Pois}(\lambda_{2j}+\lambda_{3j})) &  \;\;\text{w.p.} \;\;\;\; p_{2j}, \\
     \hbox{BivPois}(\lambda_{1j}, \lambda_{2j},\lambda_{3j})  & \;\;\text{w.p.}  \;\;\;\;  p_{3j}.
  \end{array}
\right.
\label{BivZIP}
\end{eqnarray}
where ``w.p." denotes a shorthand for ``with probability" and $p_{0j}=(1-p_{1j}-p_{2j}-p_{3j})$, denotes the
probability that observations follow a bivariate Poisson
distribution with joint probability mass function defined in
$(\ref{eq:pmf})$. Henceforth, we say that $(Y_{1j},Y_{2j})^\prime\sim\hbox{BivZIP}(\lambda_{1j},\,
\lambda_{2j},\,\lambda_{3j}, p_{1j},p_{2j}, p_{3j})$ if $(Y_{1j},Y_{2j})^\prime$ follows the distribution defined by (\ref{BivZIP}).  For a comprehensive discussion on bivariate zero-inflated
Poisson models (and extensions to multivariate cases) see \citet{li1999multivariate} and the references therein.

Semiparametric models provide an extremely versatile tool for describing nonlinear relationships and, thus, have become increasingly more prevalent among various scientific disciplines \citep[e.g.,][]{fahrmeir2006structured, lam2006semiparametric, chiogna2007semiparametric, dagne2010bayesian, musio2010bayesian, liunonparametric}.  Nevertheless, semiparametric approaches to modeling bivariate zero-inflated count data, although providing a natural framework for modeling many nonlinear ecological and environmental phenomena, remains undeveloped.  Therefore, even though we are motivated by a specific application (relative abundance for multiple species), the methodology proposed here is of independent interest.

The semiparametric modeling framework we consider uses general spline based nonparametric regression for univariate predictors \citep{ruppert2003semiparametric, ruppert2009semiparametric}.   More specifically, we consider multinomial logit models for the mixture probabilities \citep[e.g., see][]{fahrmeir2001multivariate} and semiparametric regression models for the logarithm of the latent Poisson intensity parameters.  In general, the models we propose for the intensity parameters and mixture probabilities can be expressed as
\begin{eqnarray}
\log(\lambda_{\ell j})&=&\beta_{\ell, 0}+\beta_{\ell, 1}x_{1j}+\cdots+\beta_{\ell, m_{\ell 1}-1}x_{m_{\ell 1}-1,j}+f_{\ell, m_{\ell 1}}(x_{m_{\ell 1},j})+f_{\ell, m_{\ell 1}+1}(x_{m_{\ell 1}+1,j})\nonumber\\
&&+\cdots+f_{\ell, m_{\ell 2}}(x_{m_{\ell 2},j})+\varepsilon_{\ell j} \label{loglin}\\
\log(p_{rj}/p_{0j})&=&\gamma_{r0}+\gamma_{r1}x_{1j}+\gamma_{r2}x_{2j}+\cdots+\gamma_{r{q_r}}x_{{q_r}j},\label{logitp}
\end{eqnarray}
where $\ell=1,2,3$, $r=1,2,3$ and $\varepsilon_{\ell j}$ are assumed to be i.i.d. $N(0,\sigma^2_{\varepsilon_\ell})$ and the intercepts can be considered random effects that help account for uncertainties arising from
sampling errors and covariates potentially missing from the
analysis.   Additionally, each function $f_{\ell i}(\cdot)$ is an unknown smooth function that is assumed to be approximated sufficiently well using a penalized spline.  In particular, the smooth functions $f_{\ell i}(\cdot)$ in (\ref{loglin}) are based on thin-plate splines and can be generically
written in the form
\begin{eqnarray}
f(x| \cdot)= \beta x+ \sum_{k=1}^K b_{k}
  |x-\kappa_k|^3,\label{semipar}
\end{eqnarray}
where $b_k\sim N(0, \sigma_u^2)$ and $\kappa_1\le\kappa_2\le\cdots\le\kappa_{K}$ denote fixed knot points.  Here, we focus on thin plate splines because of their good numerical properties and note that other orthogonal basis functions could also be used in this context.

For ease of exposition, we temporarily suppress the dependence on $\ell$ and consider a special case of (\ref{loglin}). In  particular, let
\begin{eqnarray}
\tilde{\lambda}_j=\log(\lambda_{j})=\beta_{0}+\beta_{1}x_{1j}+ f(x_{2,j})+\varepsilon_{j}
\end{eqnarray}
and
\begin{eqnarray*}
\begin{array}{cc}
\mbf{X}=\left[1\,\,\, x_{1j}\,\,\, x_{2j}\right]_{1\le j\le
n};&\mbf{Z}_{K}=[\underset{1\le k\le
{K}}{|x_{2j}-\kappa_k|^3}]_{1\le j\le n}.
\end{array}
\end{eqnarray*}
Further, let
\begin{eqnarray*}
\mbv{\Omega}_K=[\underset{1\le k,k'\le K}{|\kappa_k-\kappa_{k'}|^3}],
\end{eqnarray*}
and $\mbv{\tilde{\lambda}}=(\tilde{\lambda}_1,\ldots,\tilde{\lambda}_n)'$, then the penalized spline regression is obtained by minimizing
\begin{eqnarray}
\frac{1}{\sigma^2_\varepsilon}||\mbv{\tilde{\lambda}}-\mbf{X}\mbv{\beta}-\mbf{Z}_K\mbf{b}||^2 + \frac{1}{\delta\sigma^2_\varepsilon}\mbf{b}'\mbv{\Omega}_K\mbf{b},\label{matrixnotation}
\end{eqnarray}
where $\mbv{\beta}=(\beta_0,\beta_1,\beta_2)'$, $\mbf{b}=(b_1,\ldots,b_k)'$ and $\delta$ corresponds to a fixed \textit{smoothing} parameter (i.e., penalty parameter).  Additionally, let $\mbv{\beta}$ be fixed and $\mbf{b}$ random, with $\hbox{E}(\mbf{b})=\mbf{0}$, $\hbox{cov}(\mbf{b})=\sigma^2_u\mbv{\Omega}_k^{-1}$, where $\sigma^2_u=\delta\sigma^2_\varepsilon$.  As long as $(\mbf{b}',\,\mbv{\varepsilon}')'$ is normally distributed, where $\mbv{\varepsilon}=(\varepsilon_1,\varepsilon_2,\ldots,\varepsilon_n)'$, with $\mbf{b}$ and $\mbv{\varepsilon}$ independent, one can obtain an equivalent mixed model representation of the penalized spline \citep{brumback1999variable}; see \citet{crainiceanu2005bayesian}, \citet{gimenez2006semiparametric}, \citet{holan2008semiparametric} and the references therein for complete details.  Specifically, the P-spline representation of the generalized linear mixed model (GLMM) is given by
\begin{eqnarray}
\mbv{\tilde{\lambda}}=\mbf{X}\mbv{\beta}+\mbf{Z_K}\mbf{b}+\mbv{\varepsilon},\label{LMM}
\end{eqnarray}
with
\begin{eqnarray*}
\hbox{cov}\left(%
\begin{array}{c}
  \mbf{b} \\
  \mbv{\varepsilon} \\
\end{array}%
\right)=
\left(%
\begin{array}{cc}
  \sigma^2_u\mbv{\Omega}_K^{-1} & 0 \\
  0 & \sigma^2_\varepsilon \mbf{I}_n \\
\end{array}%
\right).
\end{eqnarray*}
Again, following \citet{crainiceanu2005bayesian}, define $\mbf{Z}=\mbf{Z}_K\mbv{\Omega}_K^{-1/2}$ and $\mbf{b}=\mbv{\Omega}_K^{-1/2}\mbf{u}$ the equivalent P-spline model, for the log-intensity parameters, in the form of a GLMM can be expressed as
\begin{eqnarray}
\mbv{\tilde{\lambda}}=\mbf{X}\mbv{\beta}+\mbf{Z}\mbf{u}+\mbv{\varepsilon},\label{LMM_reparam}
\end{eqnarray}
with
\begin{eqnarray*}
\hbox{cov}\left(%
\begin{array}{c}
  \mbf{u} \\
  \mbv{\varepsilon} \\
\end{array}%
\right)=
\left(%
\begin{array}{cc}
  \sigma^2_u\mbf{I}_K & 0 \\
  0 & \sigma^2_\varepsilon \mbf{I}_n \\
\end{array}%
\right).
\end{eqnarray*}

Although it is possible to write out (\ref{loglin}) explicitly in terms of its equivalent mixed model formulation, we do not pursue this general exposition for the sake of brevity.  Additionally, inclusion of spatial effects is straightforward and simply amounts to including a bivariate radial basis smoother (or basis corresponding to a proper spatial covariance) in (\ref{loglin}) (i.e., a \textit{geoadditive model}); see \citet{kammann2003geoadditive} and \citet{holan2008semiparametric} for complete details.  Discussion of specific models in the context of our motivating example is deferred until Section~\ref{application}.

One method of estimating this model is known as {\it penalized
quasilikelihood} (PQL) and constitutes an approximation to the
full likelihood \citep{breslow1993approximate, ruppert2003semiparametric}.  Another method for fitting generalized linear mixed
models is to adopt  a Bayesian approach and use Markov chain
Monte Carlo \citep{robert2004monte}, which is the direction
we pursue; see \citet{zhao2006general} for a comprehensive discussion.

Using (\ref{BivZIP}) we propose hierarchical Bayesian semiparametric  BivZIP models.  Let $[y|x]$ and $[x]$ denote the conditional distribution of $y$ given $x$ and the unconditional distribution of $x$, respectively.
Following \citet{wikle2003hierarchicala}, and assuming conditional independence, the joint posterior distribution of the catch intensity, zero inflation probability and parameters, conditional on the data can be obtained using Bayes theorem.

To completely specify a Bayesian semiparametric
BivZIP model, we need to provide prior distributions for all
parameters.  Specifically,  for
$\ell=1,2,3$ and $\mbf{u}_{\ell i}=(u_{\ell i1},\ldots,u_{\ell iK_{\ell i}})'$, we choose the prior densities as $\beta_{\ell
i}\sim N(0,\sigma^2_{\beta_{\ell i}})$,
$u_{\ell ik}\sim N(0,\sigma^2_{u_{\ell i}})$ $(k=1,\ldots,K_{\ell_i})$,
and $\gamma_{r l}\sim N(0,\sigma^2_{\gamma_{r l}})$ with the hyperparameters $\sigma^2_{\beta_{\ell i}}$ and
$\sigma^2_{\gamma_{r l}}$ assumed known and chosen by the
practitioner. In addition, the prior for $\sigma^{2}_{u_{\ell i}}$ is chosen such that
$\sigma^{2}_{u_{\ell i}}\sim\hbox{IG}(c,d)$, where $\hbox{IG}(c,d)$  corresponds to an Inverse Gamma distribution with shape parameter $c$ and scale parameter $d$.

As previously noted, in order to fit our model, we take a Bayesian MCMC
approach.  Nevertheless, many of the full conditional distributions needed to
carry out the estimation will not be of standard form and so
more sophisticated MCMC methods will be required such as
Metropolis within Gibbs \citep[see][for a comprehensive overview]{robert2004monte}.  Full details surrounding the full conditionals and MCMC algorithm are provided in the Appendix.

\section{Application to Modeling Catch Per Unit Area}\label{application}
\subsection{Model of abundance}
In order to model CPUA for multiple
species in the Missouri River benthic fish study we use the semiparametric BivZIP models proposed in Section~\ref{BVZIP}.
As alluded to, these models attempt to simultaneously account for both sources
of zeros present in our data, namely ``sampling" and
``structural" zeros, by using indicator variables corresponding
to gears, macrohabitats and segments as covariates for
describing the zero-inflation probability.  The covariates
include: four different gears, including benthic trawl (BT), drifting trammel net (DTN),
beach seine (BS), and electrofishing (EF), where EF is considered as a baseline (i.e.,
set to zero); four macrohabitats, including tributary mouth
(TRM), secondary channel-connected (SCC), secondary channel
not-connected (SCN), and Bend with TRM taken as the baseline, and three years (1996-1998)
with 1998 set as baseline. Of the existing 15 segments in this study, due to extreme sparsity ($\ge 95\%$ zeros), only ten segments (3, 5, 9,
15, 17, 19, 22, 23, 25, 27) were included in the analysis with
segment 27 set as the baseline.  Another variable used in the model
describes the substrate composition (proportion of sand,
gravel, and silt), where proportion silt is omitted from the
model (since the proportions of silt, sand, and gravel are constrained to sum to
one). Note that, aside from interpretation, the choice of baseline is arbitrary and has no
effect on the analysis.

Additionally, similar to \citet{wildhaber2011},  combined with extensive explanatory analysis, several
continuous variables are considered in the model, including, depth, water temperature, conductivity,
turbidity (log turbidity) and velocity. Ultimately, water
temperature, depth, conductivity and log turbidity are candidates to be modeled nonlinearly, whereas velocity is excluded from
the model due to high correlation with depth.

The fish counts for channel catfish (Species 1) and common carp
(Species 2) are assumed to follow a BivZIP distribution, as
described in Section \ref{BVZIP}.  More specifically, we formulate a semiparametric
hierarchical Bayesian model starting with the assumption that $(Y_{1j}, Y_{2j})' \sim\hbox{BivZIP}(a_j\lambda_{1j},a_j\lambda_{2j},a_j\lambda_{3j},p_{1j},p_{2j},p_{3j})$, where $a_j$ accounts for the different areas covered by gears
(or ``level of effort'') involved in measurement $j$
($j=1,\ldots,n$). The ``normalization" by level of effort is important and
allows us to model catch per unit area data obtained by
multiple gears. Without this normalization, the models would only apply to the case of single gears.  Now, for $\ell = 1,2, 3$, corresponding to Species 1, Species 2, and the common process, and for $r=1,2,3$ let
\begin{eqnarray}
\log(\lambda_{\ell j})&=&\beta_{\ell,1}\texttt{gear}_j+\beta_{\ell, 2} \texttt{segment}_j
+ \beta_{\ell, 3}\texttt{macrohab}_j+\beta_{\ell, 4}\texttt{year}_j
+ \beta_{\ell, 5}\texttt{substrate}_j\nonumber\\
&+&  f_{\ell, 6}(\texttt{depth}_j)+f_{\ell, 7}(\texttt{temp}_j)+f_{\ell, 8}(\texttt{lturb}_j)
+f_{\ell, 9}(\texttt{conduct}_j)+\varepsilon_{\ell j}\label{eqn:fishmvpoiloglin}, \\
\hbox{log}(p_{rj}/p_{0j})&=&\gamma_{r0}+\gamma_{r1}\texttt{gear}_j
+\gamma_{r2}\texttt{segment}_j +  \gamma_{r3}\texttt{macrohab}_j
+\gamma_{r4}\texttt{year}_j \nonumber\\
&+&\gamma_{r5}\texttt{substrate}_j.\label{eqn:fishmvpoilogit}
\end{eqnarray}
Note that, taken together,  (\ref{eqn:fishmvpoiloglin}) and (\ref{eqn:fishmvpoilogit}) are considered the ``full'' model; however, in general, each of the log-intensity models may have
different covariates and semiparametric specifications.  As described below, several variants of this model are considered in our analysis and are detailed in Table~\ref{tab:DIC}.

Although in (\ref{eqn:fishmvpoiloglin}) and (\ref{eqn:fishmvpoilogit}) it is relatively straightforward to include interaction terms for different subsets of the covariates \citep[see][]{ruppert2003semiparametric}; in practice, this requires careful monitoring.  Specifically, in the zero-inflated case, sparsity is more likely to occur within a given level of an interaction term, which may cause problems with estimation due to lack of information.  Based on sparsity considerations and the fact that, in our case, models without interaction terms exhibited lower deviance information criteria (DIC) values \citep{spiegelhalter2002bayesian}, the models presented here only consider main effects.

The smooth functions $f_{\ell i
}(\cdot)$ in (\ref{eqn:fishmvpoiloglin}) are based on
thin-plate splines, as defined by (\ref{semipar}), with 20 knot points equally spaced in the covariate domain. Note that several of the
covariates are
categorical, which complicates interpretation of the
intercepts. The coefficients of the model
corresponding to a specific level of a categorical variable can be roughly
interpreted as the log ``mean" fish CPUA in that specific level, relative to the
baseline level (the level set to zero), while holding all other
variables fixed.

For $\ell=1,2,3$, we define relatively noninformative
prior densities for the unknown parameters $\beta_{\ell i }$ and $\gamma_{r i}$ as
$N(0,100)$, and for $u_{\ell i }$ parameters we
define the prior density as $N(0,\sigma^2_{u_{\ell
i}})$. Additionally, the prior for $\sigma^{2}_{u_{\ell i }}$ is chosen as $\sigma^{2}_{u_{\ell i }}\sim\hbox{IG}(c,d)$
with $c$ and $d$ specified such that the prior mean and variance are equal to .001 and 100, respectively.
Note that, in all cases, the priors chosen are vague but proper and thus, we maintain
propriety of the posterior while imparting little impact on the analysis.

The model we propose for this application is well suited, since the species are naturally linked
together for each observation. Specifically, each
observation defines a specific gear deployment over which the count for both species is
recorded.  In particular, these species are biologically related and
thus it is expected that the CPUA for each observation, as a
function of macrohabitat, year, gear (i.e., type of net) used to
catch the fish and segment will be correlated.
Also, note that we do not consider the problem of estimating
true abundance and probability of detection in our analysis, instead, we limit our work to modeling relative abundance due to lack of
information on gear efficiencies (i.e., detection probability)
and repeated sampling. However, there are approaches available
based on repeated sampling (e.g., capture-recapture) to
estimate true abundance and detection probabilities (see \citet{royle2008hierarchical} for a complete discussion) and can be implemented in a similar manner to our proposed modeling framework.

The MCMC computations were implemented using Gibbs
and Metropolis-Hastings within Gibbs sampling algorithms (see the Appendix). A total of 120,000 MCMC
realizations were obtained with 20,000 iterations discarded for burn-in.  Subsequently, every fifth sample was kept for
inference, resulting in 20,000 iterations total. Convergence of the MCMC
chains was verified through visual inspection of trace plots of
the sample chains.

\subsection{Model Selection}
Aside from models having interaction terms, we compared the performance of seven different models using DIC.  Specifically, the models considered differed only in the form of their
continuous covariates for log(turbidity), depth, water
temperature, and conductivity (e.g., whether these covariates entered into the model linearly or nonlinearly). These
models were selected mainly based on subject matter considerations as well as exploratory data analysis.  Table~\ref{tab:DIC} provides a comprehensive overview of the models being compared.

Selection and inference based on these models helps foster a better understanding of specific linear or
nonlinear effects that the covariates have on the common and individual Poisson log-intensities. Model 1 (M1) contains nonlinear specifications for all
of the continuous covariates in all three log-intensity models (i.e., M1 is the
``full'' model).  Conversely, Model 2 (M2) has nonlinear specifications for
log(turbidity) and depth in the individual log-intensity models and has all linear
covariates for the common log-intensity model. That is, this model
assumes that the common effects of these two covariates are
linear and the nonlinear effects only arise for the individual
processes. Model 3 (M3) is similar to M2;
however, the linear terms for water temperature and
conductivity which were not statistically significant in M2, were excluded. Model 4 (M4) is similar to M3; however, all the variables that were not statistically
significant in M3 were excluded.  The variables that were excluded for individual log-intensity models are year and substrates (sand and gravel), whereas for the common log-intensity process the variables are substrates, log(turbidity), depth, water temperature and
conductivity.  Model 5 (M5) is nonlinear in all of the
continuous covariates for the individual log-intensity models and linear
in all of the covariates for the common log-intensity model. Model 6 (M6) is nonlinear for all
continuous variables for the common log-intensity model
and has linear covariates for the individual log-intensity models.
Finally, Model 7 (M7) has linear covariates for all
continuous variables in all of the log-intensity models.  Our selection criteria favor the
most parsimonious model with the smallest DIC value.

\subsection{Results}\label{results}
Based on the DIC values (see Table~\ref{tab:DIC}), M3 is
considered to be the ``best'' model among all the models
considered; thus, we only discuss the results of M3 here.   Recall, M3 does not
include water temperature or conductivity and the only
nonlinear effects,log(turbidity) and depth,  arise only in the
individual log-intensity models.

Posterior means and 95\% credible intervals (CI) for the
log-intensity models are given in Table \ref{tab:loglin}. A
covariate is considered to have a ``significant" effect if its
credible interval does not include 0. Note that for indicator
variables corresponding to the levels of categorical variables,
results are based on comparison of the variable relative to the baseline. Our model results for the
log-intensity models (Table~\ref{tab:loglin}) corroborate
the univariate results of the previous analyses conducted by \citet{berry:wildhaber:young:2005} and \citet{wildhaber2011} . However, our
model also provides results for the common process; e.g., information about
common habitat usage for both species. In contrast, the previous analyses of \citet{berry:wildhaber:young:2005}, \citet{arab2008zero} and \citet{wildhaber2011} are incapable of delivering this information in a concise manner.

Based on the results
for the common process, the mean CPUA for both species is
higher in BEND macrohabitats compared to TRM (tributary mouth). Also, mean CPUA
for both species increased in year 1998 compared to year 1996.
Electrofishing is best when considering both species (although
not the best gear for sampling channel catfish only). Importantly, this
information provides fisheries biologists with better
understanding about the common habitat usage and
distribution of the two species in the Missouri River.

\begin{figure}
\begin{center}
 \subfigure[]{\includegraphics[width=3in,height=3in]{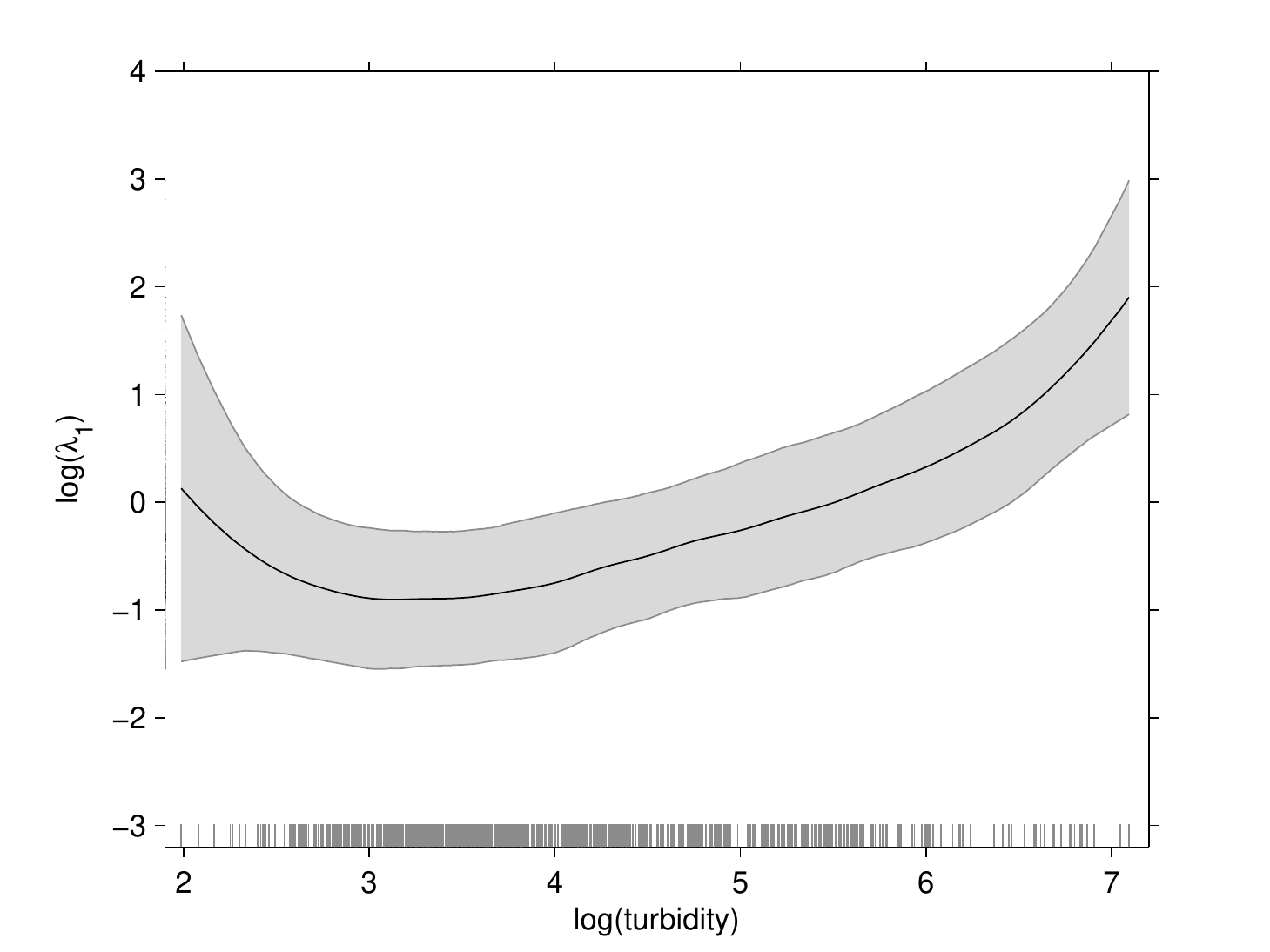}}
  \subfigure[]{\includegraphics[width=3in,height=3in]{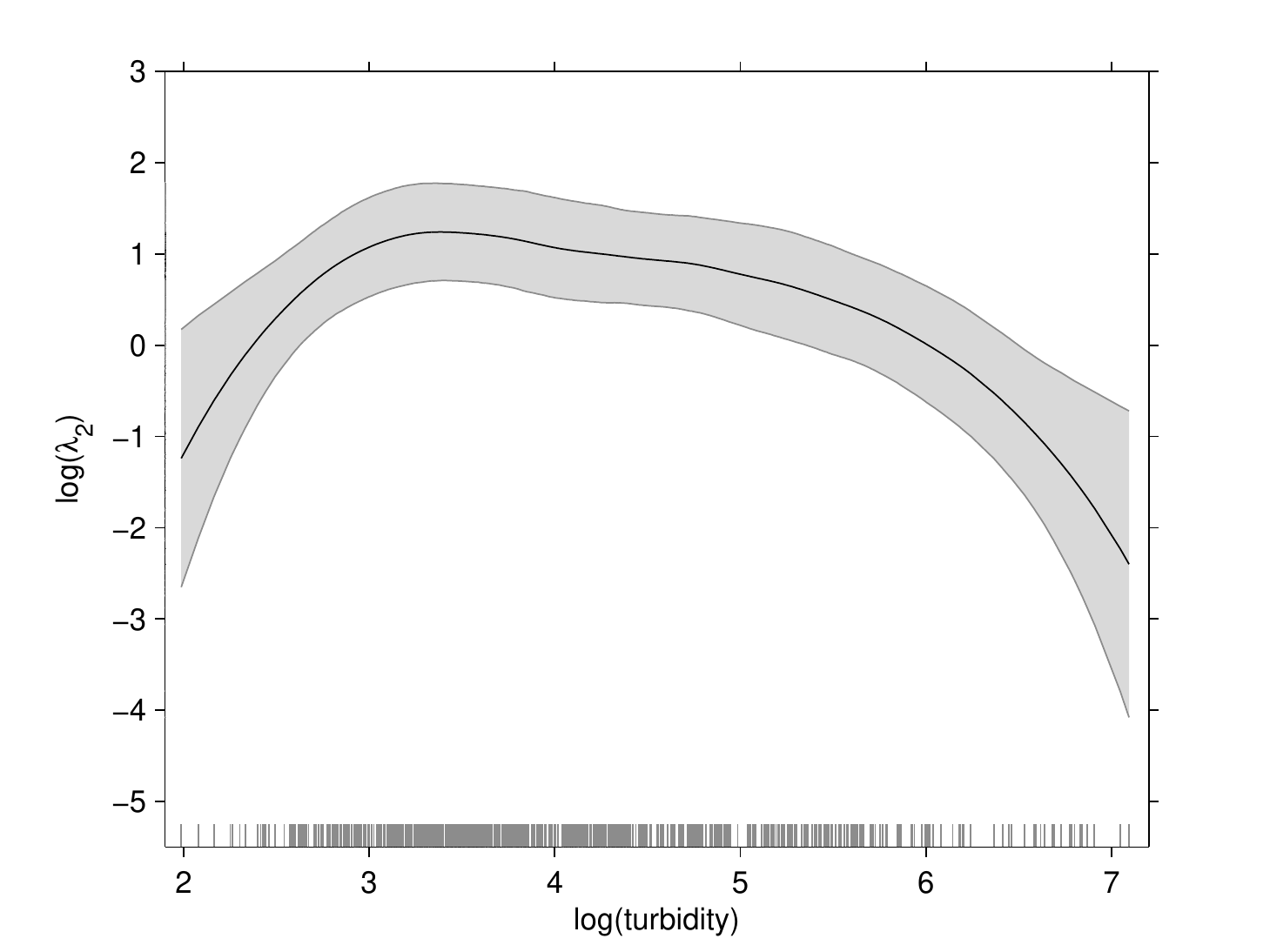}}\\
\subfigure[]{\includegraphics[width=3in,height=3in]{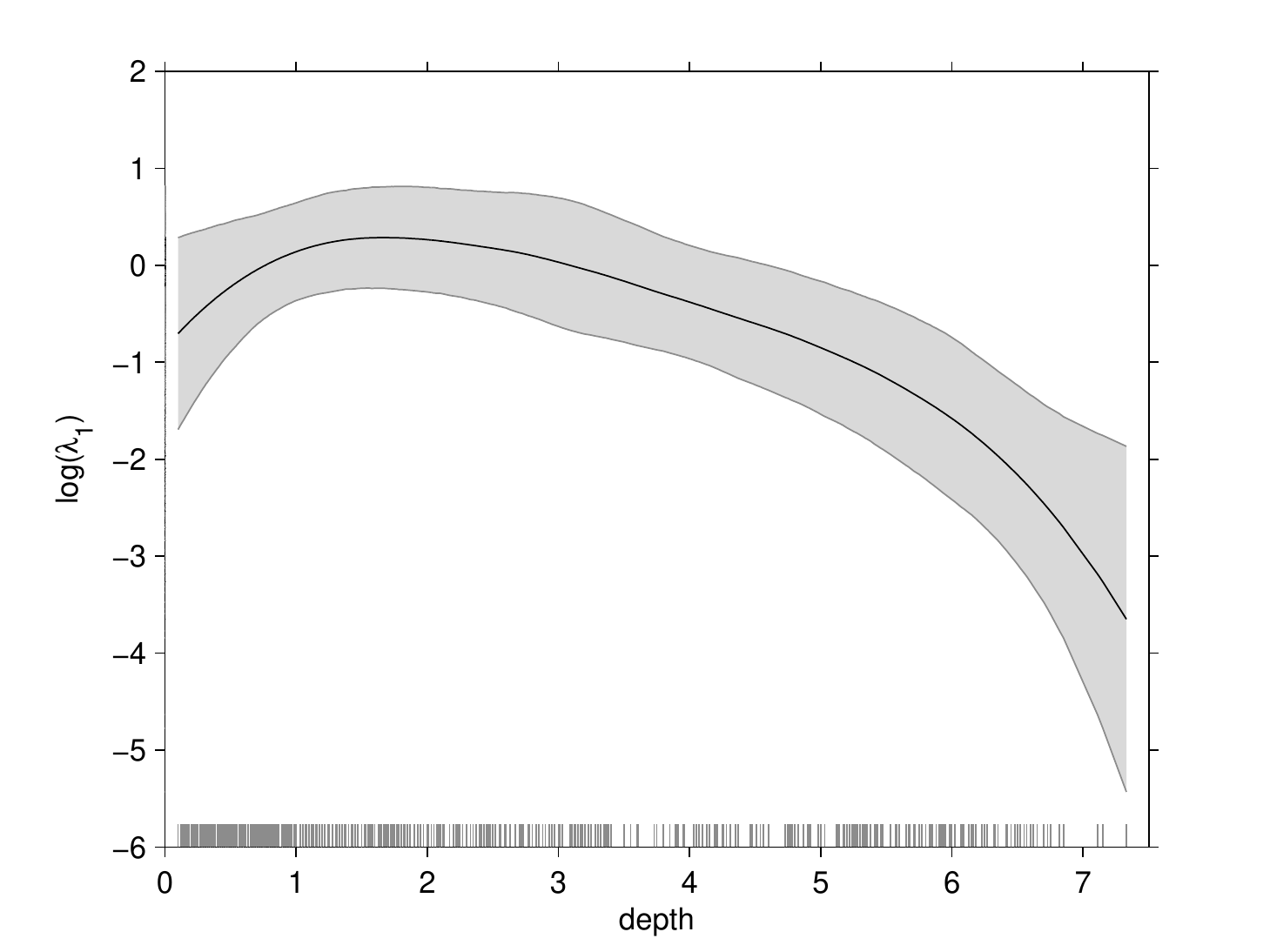}}
  \subfigure[]{\includegraphics[width=3in,height=3in]{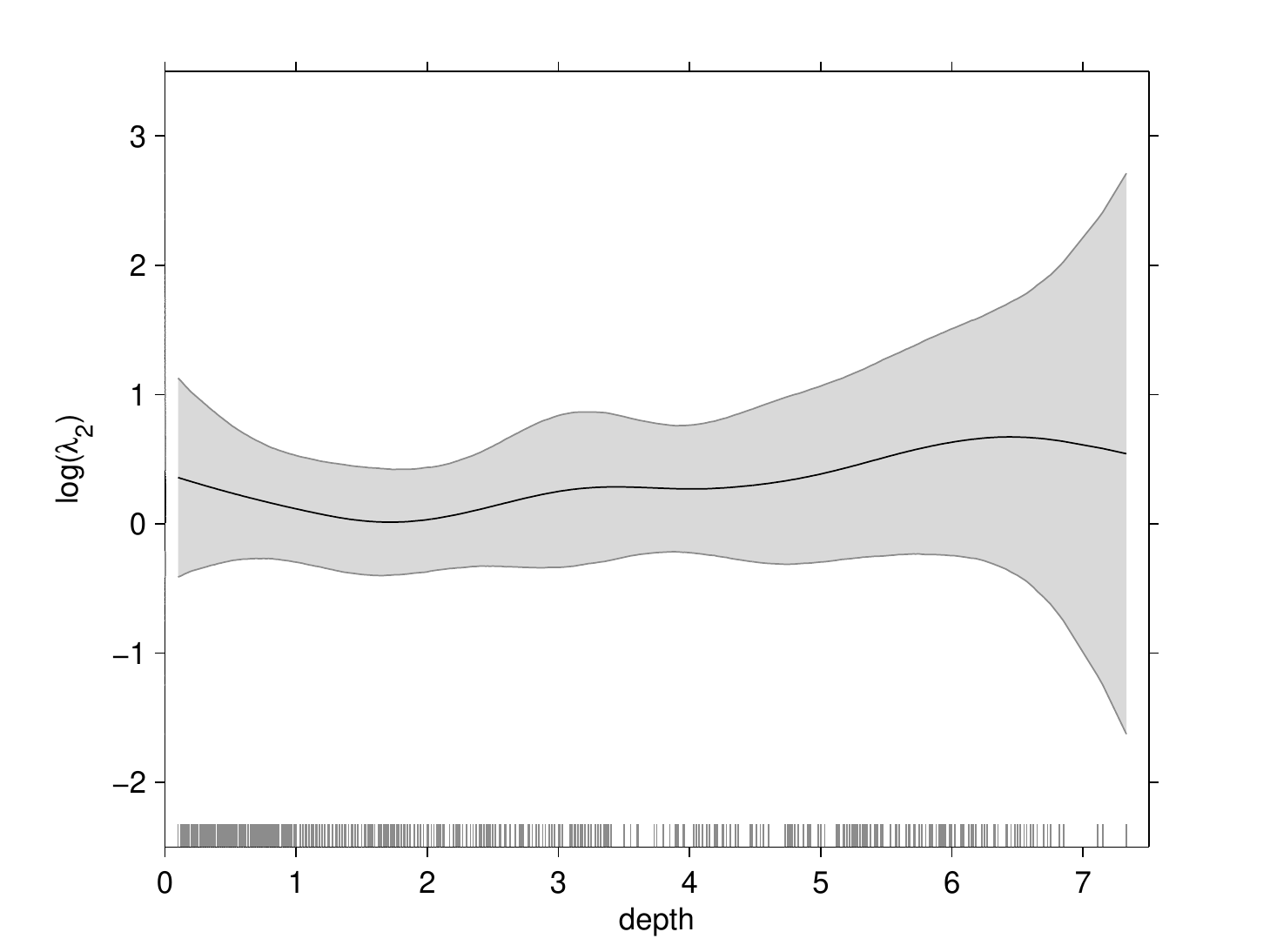}}\\
  \caption{ \baselineskip=10pt Posterior mean for the nonlinear functions in M3 (solid line) and point-wise 95\% CIs (shaded area) for both species: (a) log(turbidity) (channel catfish), (b) log(turbidity) (common carp),
(c) depth (channel catfish), and (d) depth (common carp).}
  \label{fig:semipar1}
\end{center}
\end{figure}

Of particular interest in this analysis is the potential
nonlinear effect of log(turbidity) and depth on CPUA for
these two species. As previously noted, channel catfish and common carp are both generalist species and thus tend to
exhibit similar habitat selection. Our results corroborate the findings of \citet{wildhaber2011}  but provide more detailed information about
the effects of depth and turbidity on the relative abundance of
these two species  (Figure~\ref{fig:semipar1}). Specifically, a nonlinear relationship
between log(turbidity) and the BivZIP log-intensity parameters is present for both
species (Figure~\ref{fig:semipar1} (a) and (b)). In particular,
these figures demonstrate that the mean CPUA of channel catfish
increases with higher turbidity whereas the mean CPUA of common
carp was highest around mid-range levels of turbidity but
decreased for higher turbidity values. Similarly, depth is
nonlinearly related to the BivZIP log-intensity parameters for channel catfish with peaks
around lower depth levels (1 to 3 meters; Figure~\ref{fig:semipar1} (c)). However, given the relatively higher
uncertainty in the nonlinear plot of depth for common carp
(Figure~\ref{fig:semipar1} (d)), there does not seem to be
strong evidence to support a potential nonlinear effect of depth
on the log-intensity parameter for common carp.

Finally, several factors significantly impact the
zero-inflation probabilities for both species.  For the sake of
brevity, discussion is limited to the most salient results, with a comprehensive list provided in Table~\ref{tab:logp}. The
multinomial logit model on probabilities corresponds to the log ratio of
$p_1$, $p_2$, and $p_3$ relative to the baseline
probability $p_0$, which is the probability of observing zero
values for both species. For the model corresponding
to $p_1$, the probability that channel catfish data is
non-zero and common carp is zero, the only statistically significant (positive) effect, relative to the baseline (segment 27),
is segment 15,  which is also significantly higher than segments 19 and 25, relative to the baseline.  For the model corresponding to
$p_2$, the probability that channel catfish observation
is zero and common carp is non-zero, the only significant
effect, relative to the baseline (TRM), is BEND, which is significantly less than zero.
Finally, for the model corresponding to $p_3$,
the probability that observations for both species are
non-zero, relative to the baseline, segment 9 is significantly greater than zero and significantly higher than segments 17 and 19, whereas drifting trammel net is significantly lower than the baseline (electrofishing) and all other gears (relative to the baseline).

It is important to stress that these findings are based on a
bivariate model and provide
information that was unobtainable in previous
analyses of these data. The environmental variables were not
included in the final multinomial logit models presented for the probabilities,
due to the fact that  a ``full" model including these variables (not shown) exhibited no significant effects (for these variables). These
findings provide useful information surrounding the presence and absence
of both species based on environmental attributes of the river
as well as sampling procedures (i.e.,  gears).  Ultimately this information is crucial to fisheries biologists interested in optimizing management efforts and designing future monitoring programs.

\section{Conclusion}\label{sec:concl}
Ecological and environmental processes often produce count data
having a high proportion of zeros.  Typically these processes
are related to externally measured explanatory variables
through complex nonlinear relationships.  Furthermore, when
developing models of species abundance it is often advantageous
to allow correlation among the response variable for multiple
populations under consideration.  We propose an
extremely flexible hierarchical Bayesian semiparametric BivZIP model and note that
this model can be utilized across a diverse range of applications
including those outside the natural sciences. Moreover, our
approach represents the first attempt at semiparametric
modeling of bivariate zero-inflated counts.  The effectiveness of our approach is demonstrated through a
motivating application to modeling fish catch per unit area
based on observations from two closely relates species of
benthic fish in the Missouri River.

The example presented here illustrates
the utility of our model for drawing inference on complex
ecological data.  In particular, our model accommodates
inherent nonlinear relationships, while allowing for the high
proportion of zeros in the data.  In several instances we are
able to provide inference where traditional modeling approaches
were unsuccessful.  For instance,  despite the fact that
each species exhibited a large percentage of zero catches, we were able
to find a significant effect due to gears.  Furthermore, our approach was able to show, with
statistical significance, higher catch for high log(turbidity)
values and higher catch for low to mid-range depth values for channel catfish
and low to mid-range log(turbidity) values for common carp.

Our model borrows strength across correlated species,
which is fundamentally important for successful modeling of
abundance of multiple species. One
potential practical limitation of  any bivariate (or multivariate) approach, including ours, is that by
pairing data for two or more species the number of cases that
may have to be excluded from the model due to extreme sparsity or missing data  increases.
Nonetheless, in our analysis, we are still able to exclude fewer observations than would be necessary using standard statistical methods.  In addition, using our approach, missing data can be readily dealt with using data
imputation techniques, which are particularly easy to implement
in the Bayesian framework.  Finally, our method can be adapted to further avoid excluding data arising from sparsity due to low catch rates.  In this case, science based prior distributions or data from previous and/or similar studies may be readily incorporated into the model hierarchy.

%
%
%
%
%
%
%
%
%
%\paragraph{Acknowledgments} We would like to thank an anonymous
%referee for constructive comments that led to improvement of the
%original manuscript.
%

\section*{Appendix: Conditional Distributions for MCMC}
\begin{appendix}
\Appendix    % This makes the section title start with Appendix!
\renewcommand{\theequation}{A.\arabic{equation}}
\setcounter{equation}{0}

%\textbf{***Ali - Please put the full conditionals and sampling algorithm here.***}
%\subsection*{Appendix: Full-Conditionals}

In this section, we describe the MCMC algorithm for Model 3 (M3) and note that the migration to other models is analogous. Recall,
M3 is given by
\begin{eqnarray}
\log(\lambda_{\ell j})&=&\beta_{\ell,1}\texttt{gear}_j+\beta_{\ell, 2} \texttt{segment}_j
+ \beta_{\ell, 3}\texttt{macrohab}_j+\beta_{\ell, 4}\texttt{year}_j
+ \beta_{\ell, 5}\texttt{substrate}_j\notag\\
&+&  f_{\ell, 6}(\texttt{depth}_j)+f_{\ell, 7}(\texttt{lturb}_j)
+\varepsilon_{\ell j}\label{eqn:fishmvpoiloglinapp}, \\
\hbox{log}(p_{rj}/p_{0j})&=&\gamma_{r0}+\gamma_{r1}\texttt{gear}_j
+\gamma_{r2}\texttt{segment}_j +  \gamma_{r3}\texttt{macrohab}_j
+\gamma_{r4}\texttt{year}_j \notag\\
&+&\gamma_{r5}\texttt{substrate}_j,\label{eqn:fishmvpoilogitapp}
\end{eqnarray}

\noindent for $\ell=1,2,3$, $r=1,2,3$, and $j=1,\ldots,n$. Using Bayes' theorem, and assuming conditional independence, the posterior distribution of the processes and parameters given
the observations can be expressed as
\begin{align}
  [ \mbv{\tilde{\lambda}}_{1}, & \mbv{\tilde{\lambda}}_{2}, \mbv{\tilde{\lambda}}_{3},\textbf{B}_{1},\textbf{B}_{2},\textbf{B}_{3},\mbv{\gamma}_{1}, \mbv{\gamma}_{2},\mbv{\gamma}_{3}, \textbf{u}_{1, \texttt{lturb}},\textbf{u}_{1, \texttt{depth}}, \textbf{u}_{2, \texttt{lturb}}, \textbf{u}_{2, \texttt{depth}},
 \sigma^2_{u_1,\texttt{lturb}}, \sigma^2_{u_1,\texttt{depth}},\sigma^2_{u_2,\texttt{lturb}}, \sigma^2_{u_2,\texttt{depth}},
 \notag \\  & \sigma^2_{\varepsilon_1}, \sigma^2_{\varepsilon_2},\sigma^2_{\varepsilon_3}| \textbf{Y}_{1},
 \textbf{Y}_{2}] \propto  [\textbf Y_{1},\textbf Y_{2}|\mbv{\tilde{\lambda}}_{1},  \mbv{\tilde{\lambda}}_{2}, \mbv{\tilde{\lambda}}_{3},\mbv{\gamma_1}, \mbv{\gamma_2}, \mbv{\gamma_3}][\mbv{\tilde{\lambda}}_{1}|\textbf{B}_{1},\textbf{u}_{1, \texttt{lturb}},
 \textbf{u}_{1, \texttt{depth}},
\sigma^2_{\varepsilon_1},\sigma^2_{u_1,\texttt{lturb}},\sigma^2_{u_1,\texttt{depth}}]\notag \\& \times [\mbv{\tilde{\lambda}}_{2}|\textbf{B}_{2},\textbf{u}_{2, \texttt{lturb}},\textbf{u}_{2, \texttt{depth}},
\sigma^2_{\varepsilon_2},\sigma^2_{u_2,\texttt{lturb}},\sigma^2_{u_2,\texttt{depth}} ][\mbv{\tilde{\lambda}}_{3}|\textbf{B}_{3}, \sigma^2_{\varepsilon_3}][\mbv \gamma_{1}][\mbv \gamma_{2}] [\mbv \gamma_{3}][\textbf{B}_{1}][\textbf{B}_{2}][\textbf{B}_{3}][\textbf{u}_{1}][\textbf{u}_{2}] [\sigma^2_{\varepsilon_1}]\notag\\&
 \times
[\sigma^2_{\varepsilon_2}][\sigma^2_{\varepsilon_3}][\sigma^2_{u_1,\texttt{lturb}}][\sigma^2_{u_1,\texttt{depth}}][\sigma^2_{u_2,\texttt{lturb}}][\sigma^2_{u_2,\texttt{depth}}], \notag
\end{align}

\noindent where $\textbf{B}_{\ell}$ denotes the vector of coefficients associated with the linear effects in each log-intensity model.

\noindent This complicated posterior distribution can be numerically
evaluated using MCMC methods and, in particular, Gibbs sampling
based on full-conditional distributions of the unknown processes and
parameters.  Also, Metropolis-Hastings (M-H)  steps within the Gibbs
algorithm are required for simulation of $\mbv{\tilde{\lambda}}_\ell=\log(\mbv{\lambda}_\ell)$ ($\ell=1,2,3$) and
$\mbv{\gamma}_r$ ($r=1,2,3$).\\

\noindent The full-conditional distributions and sampling steps are:

\begin{itemize}
\item Generate the latent process ($\textbf{Z}_3$). For $j=1,\ldots,n$, draw each element of $\textbf{Z}_3$ (i.e., $Z_{3j}$) from a DUnif$\left\{0, \text{min}(Y_{1j}, Y_{2j})\right\}$ , then set
\begin{center}
$\textbf{Z}_1=\textbf{Y}_1-\textbf{Z}_3$ and $\textbf{Z}_2=\textbf{Y}_2-\textbf{Z}_3$.
\end{center}

  \item  $[\mbv{\tilde{\lambda}}_{\ell} | \cdot]$ \hspace{.1in}($\ell=1,2$)\\
  \text{M-H step:} \\
  \text{1. Generate a candidate $\mbv{\tilde{\lambda}}_{\ell}^{*}\thicksim N(\mbv{\tilde{\lambda}}_{\ell}^{(i-1)},\theta_\ell \textbf{I})$ at the $i$th MCMC iteration (where $\theta_\ell$ is a}\\
   \text{tuning parameter chosen such that the acceptance rate for the M-H algorithm is between }\\
   \text{20\% and 40\%), and compute the ratio}\\
  \begin{displaymath}
  \hspace{-.25in} \textbf{R}= \frac{[\textbf{Y}_{1}, \textbf{Y}_{2}|\mbv{\tilde{\lambda}}_{\ell}^{*}][\mbv{\tilde{\lambda}}_{\ell}^{*}| \mbv{\Phi}_\ell^{(i-1)},\sigma^{2,(i-1)}_{\varepsilon_\ell},\sigma^{2,(i-1)}_{u_\ell, \texttt{lturb}}, \sigma^{2,(i-1)}_{u_\ell, \texttt{depth}}]}{[\textbf{Y}_{1}, \textbf{Y}_{2}|\mbv{\tilde{\lambda}}_{\ell}^{(i-1)}][\mbv{\tilde{\lambda}}_{1\ell}^{(i-1)}| \mbv{\Phi}_\ell^{(i-1)},\sigma^{2,(i-1)}_{\varepsilon_\ell},\sigma^{2,(i-1)}_{u_\ell, \texttt{lturb}}, \sigma^{2,(i-1)}_{u_\ell, \texttt{depth}}]},\\
  \end{displaymath}

  where $\mbv{\Phi}_\ell=(\textbf{B}'_\ell, \textbf{u}_{\ell, \texttt{lturb}}', \textbf{u}_{\ell, \texttt{depth}}')'$.\\
  \vspace{.1in}
   \text{2. Set $\tilde{\lambda}_{\ell j}^{(i)}=\tilde{\lambda}_{\ell j}^{*}$ ($j=1,\ldots,n$) with probability
  min($R_j$,1); else set $\tilde{\lambda}_{\ell j}^{(i)}=\tilde{\lambda}_{\ell j}^{(i-1)}$}.
 \item $[\mbv{\tilde{\lambda}}_{3} | \cdot]$ \\
  \text{M-H step:} \\ \text{1. Generate a candidate $\mbv{\tilde{\lambda}}_{3}^{*}\thicksim N(\mbv{\tilde{\lambda}}_{3}^{(i-1)},\theta_3 \textbf{I})$ at the $i$th MCMC iteration (where $\theta_3$ is the}\\
   \text{tuning parameter chosen such that the acceptance rate for the M-H algorithm is between }\\
   \text{20\% and 40\%), and compute the ratio}\\
  \begin{displaymath}
  \textbf{R}= \frac{[\textbf{Y}_{1},\textbf{Y}_{2}|\mbv{\tilde{\lambda}}_{3}^{*}][\mbv{\tilde{\lambda}}_{3}^{*}|\textbf{B}_{3}^{(i-1)},\sigma^{2,(i-1)}_{\varepsilon_3}]}{[\textbf{Y}_{1}, \textbf{Y}_{2}|\mbv{\tilde{\lambda}}_{3}^{(i-1)}][\mbv{\tilde{\lambda}}_{3}^{(i-1)}|\textbf{B}_{3}^{(i-1)},\sigma^{2,(i-1)}_{\varepsilon_3}]}.\\
  \end{displaymath}
  \vspace{.1in}
   \text{2. Set $\tilde{\lambda}_{3j}^{(i)}=\tilde{\lambda}_{3j}^{*}$ with probability
  min($R_j$,1); else set $\tilde{\lambda}_{3j}^{(i)}=\tilde{\lambda}_{3j}^{(i-1)}$}.\\
%
% gammas for the multinom logit model
  \item  $[\mbv{\gamma}_{r} | \cdot]$ \hspace{.1in}($r=1,2,3.$)\\
  \text{M-H step:} \\ \text{1. Generate a candidate $\mbv{\gamma}_{r}^*\thicksim N(\mbv{\gamma}_r^{(i-1)},\theta_r \textbf{I})$ at the $i$th MCMC iteration, and compute}\\
  \text{the ratio}
  \begin{displaymath}
  \textbf{R}= \frac{[\textbf{Y}_1,\textbf{Y}_2| \mbv{\tilde{\lambda}}_{1}, \mbv{\tilde{\lambda}}_{2}, \mbv{\tilde{\lambda}}_{3}, \mbv{\gamma}_{r}^*][\mbv{\gamma}_{r}^*]}{[\textbf{Y}_1,\textbf{Y}_2| \mbv{\tilde{\lambda}}_{1}, \mbv{\tilde{\lambda}}_{2}, \mbv{\tilde{\lambda}}_{3}, \mbv{\gamma}_{r}^{(i-1)}][\mbv{\gamma}_{r}^{(i-1)}]}.\\
  \end{displaymath}
  \vspace{.2in}
   \text{2. Set $\gamma_{rj}^{(i)}=\gamma_{rj}^*$ ($j=1,\ldots,n$) with probability
  min($R_j$,1); else set $\gamma_{rj}^{(i)}=\gamma_{rj}^{(i-1)}$}.\\
Next, the zero-inflation probabilities are derived based on $\mbv{\gamma}_r$s at each iteration. Specifically,
$\textbf{p}_0=1/ \textbf{D}$ and $\textbf{p}_r=\exp({\textbf{X}_\gamma \mbv{\gamma}_r})/\textbf{D}$,
and $\textbf{D}=1+\sum_{r=1}^{3}\exp({\textbf{X}_\gamma \mbv{\gamma}_r}).$\\
where $\textbf{p}_0$ and $\textbf{p}_r$ ($r=1, 2, 3$) denote the vectors of probabilities and $\textbf{X}_\gamma$ denotes the design matrix for the multinomial logit regression models.

%W_k's
\item Update the regression coefficients and spline coefficients jointly:
$[\mbv{\Phi}_{\ell} | \cdot] \propto [\mbv{\tilde{\lambda}}_\ell|\mbv{\Phi}_{\ell} ][\mbv{\Phi}_{\ell}]$
%where $\mbv{\Phi}_g$ is a concatenated matrix of regression coefficients: $\mbv{\Phi}_g=(\textbf{B}_g, \textbf{u}_{g, \texttt{lturb}}, \textbf{u}_{g, \texttt{depth}})$ for the concatenated matrix of variables $\textbf{C}_g=[\textbf{X}_g \;\; \textbf{Z}_g]$ ($g=1,2$).\\
\begin{displaymath}
\mbv{\Phi}_{\ell} | \cdot \thicksim N(\textbf{A}_\ell \textbf{b}_\ell,\textbf{A}_\ell),
\end{displaymath}
with,
\begin{align}
\textbf{A}_\ell&=\left(\textbf{C}_\ell'\textbf{C}_\ell+\frac{\sigma_{\varepsilon_\ell}^{2}}{\sigma_{u_{\ell,1}}^{2}}\textbf{E}_1+ \frac{\sigma_{\varepsilon_\ell}^{2}}{\sigma_{u_{\ell,2}}^{2}}\textbf{E}_2+\frac{\sigma_{\varepsilon_\ell}^2}{\sigma_{\varepsilon_\ell,0}^2}\Sigma_{\ell}^{-1} \right)^{-1},\notag \\
\textbf{b}_\ell&=\textbf{C}_\ell'\mbv{\tilde{\lambda}}_\ell+\frac{\sigma_{\varepsilon_\ell}^2}{\sigma_{\varepsilon_\ell,0}^2}\Sigma_{\ell}^{-1}\mbv{\Phi}_{\ell,0},\notag
\end{align}

where $\textbf{E}_1=\text{diag}(\textbf{0}_m, \textbf{1}_{\kappa_\ell},\textbf{0}_{\kappa_\ell})$, $\textbf{E}_2=\text{diag}(\textbf{0}_m, \textbf{0}_{\kappa_\ell},\textbf{1}_{\kappa_\ell})$,
$\Sigma_{\ell}=\text{diag}(\textbf{1}_m, \textbf{0}_{2\times\kappa_\ell})$, $m$ denotes the number of linear effects and $\kappa_\ell \equiv 20$ is the number of knot locations. \\

% B_3
\item
$[\textbf{B}_{3} | \cdot] \propto [\mbv{\tilde{\lambda}}_3|\textbf{B}_{3} ][\textbf{B}_{3}]$\\
\begin{displaymath}
\textbf{B}_{3} | \cdot \thicksim N(\textbf A \textbf b,\textbf A),
\end{displaymath}
with,
\begin{align}
\textbf A&=\left(\textbf{X}_3'\textbf{X}_3+\frac{\sigma_{\varepsilon_3}^{2}}{\sigma_{\varepsilon_{3,0}}^{2}}\textbf{I}\right)^{-1},\notag \\
\textbf b&=\textbf{X}_3'\mbv{\tilde{\lambda}}_3+\frac{\sigma_{\varepsilon_3}^{2}}{\sigma_{\varepsilon_{3,0}}^{2}}\textbf{B}_{3,0}.\notag
\end{align}

% sigma^2
  \item $[\sigma_{\varepsilon_\ell}^2|\cdot]\propto [\mbv{\tilde{\lambda}}_{\ell}|\sigma_{\varepsilon_\ell}^2]
  [\sigma_{\varepsilon_\ell}^2]$
 \begin{displaymath}
  \sigma_{\varepsilon_\ell}^2|\cdot\thicksim \text{IG}(c_{\varepsilon_\ell},d_{\varepsilon_\ell})
\end{displaymath}
where,
\begin{align}
c_{\varepsilon_\ell}&=c_{\varepsilon_\ell,0}+ n/2, \notag \\
d_{\varepsilon_\ell}&= \left(\frac{1}{d_{\varepsilon_\ell,0}} + 0.5 (\mbv{\tilde{\lambda}}_{\ell}- \textbf{C}_{\ell}\mbv{\Phi}_{\ell})\right)^{-1}\notag,
\end{align}
where $\ell=1,2,3$ (note that for $\ell=3$, $\mbv{\Phi}_3=\textbf{B}_3$ and $\textbf{C}_3=\textbf{X}_3$), and the prior distribution for $[\sigma_{\varepsilon_\ell}^2]$ is chosen as IG($c_{\varepsilon_\ell,0}$, $d_{\varepsilon_\ell,0}$).\\
%sigma_u_k
  \item $[\sigma_{u_{\ell}}^2|\cdot]\propto [\mbv{\tilde{\lambda}}_{\ell}|\sigma_{u_\ell}^2][\sigma_{u_\ell}^2]$
\begin{displaymath}
  \sigma_{u_{\ell}}^2|\cdot\thicksim \text{IG}(c_u,d_u)
\end{displaymath}
where,
\begin{align}
c_u&=c_{u_\ell,0}+ \kappa_\ell/2, \notag \\
d_u&=\left(\frac{1}{d_{u_\ell,0}} + 0.5 \textbf{u}_\ell'\textbf{u}_\ell \right)^{-1},\notag
\end{align}
where the prior distribution for $[\sigma_{u_\ell}^2]$ is chosen as IG($c_{u_\ell,0}$, $d_{u_\ell,0}$). Note that for each of the nonlinear effects, log(turbidity) and depth, this has to be done separately; however, for brevity, we only describe the procedure for the variance component of one of the nonlinear effects.
\end{itemize}
\end{appendix}

%\clearpage\pagebreak\newpage\thispagestyle{empty}
%
\baselineskip=14pt \vskip 4mm\noindent
\bibliographystyle{jasa}
\bibliography{Biv_Semipar_ZIP_Final_ArXiv}

\begin{thebibliography}{44}
\newcommand{\enquote}[1]{``#1''}
\expandafter\ifx\csname natexlab\endcsname\relax\def\natexlab#1{#1}\fi

\bibitem[\protect\citename{Arab, }2007]{Arab2007}
Arab, A. (2007).
\newblock \enquote{Hierarchical Spatio-Temporal Models for Environmental
  Processes.}
\newblock {Unpublished Ph.D. Dissertation}, University of Missouri, Department
  of Statistics.

\bibitem[\protect\citename{Arab et~al., }2008]{arab2008zero}
Arab, A., Wildhaber, M., Wikle, C., and Gentry, C. (2008).
\newblock \enquote{{Zero-Inflated Modeling of Fish Catch per Unit Area
  Resulting from Multiple Gears: Application to Channel Catfish and Shovelnose
  Sturgeon in the Missouri River}.}
\newblock {\em North American Journal of Fisheries Management\/}, 28, 4,
  1044--1058.

\bibitem[\protect\citename{Berry et~al., }2005]{berry:wildhaber:young:2005}
Berry, C., Wildhaber, M., and Galat, D. (2005).
\newblock \enquote{Fish Distribution and Abundance. Population Structure and
  Habitat Use of Benthic Fishes Along the Missouri and Lower Yellowstone.}
\newblock Tech. rep., United States Geological Survey.

\bibitem[\protect\citename{Berry and Young, }2001]{BerryYoung2001}
Berry, C. and Young, B. (2001).
\newblock \enquote{Introduction to the Benthic Fishes Studies. Population
  Structure and Habitat Use of Benthic Fishes Along the Missouri and Lower
  Yellowstone Rivers.}
\newblock Tech. rep., United States Geological Survey.

\bibitem[\protect\citename{Breslow and Clayton, }1993]{breslow1993approximate}
Breslow, N. and Clayton, D. (1993).
\newblock \enquote{{Approximate Inference in Generalized Linear Mixed Models}.}
\newblock {\em Journal of the American Statistical Association\/}, 88, 421,
  9--25.

\bibitem[\protect\citename{Brumback et~al., }1999]{brumback1999variable}
Brumback, B., Ruppert, D., and Wand, M. (1999).
\newblock \enquote{{Variable Selection and Function Estimation in Additive
  Nonparametric Regression Using a Data-Based Prior: Comment}.}
\newblock {\em Journal of the American Statistical Association\/}, 94, 447,
  794--797.

\bibitem[\protect\citename{Chiogna and Gaetan,
  }2007]{chiogna2007semiparametric}
Chiogna, M. and Gaetan, C. (2007).
\newblock \enquote{{Semiparametric Zero-Inflated Poisson Models with
  Application to Animal Abundance Studies}.}
\newblock {\em Environmetrics\/}, 18, 3, 303--314.

\bibitem[\protect\citename{Clarke and Green, }1988]{clarke1988statistical}
Clarke, K. and Green, R. (1988).
\newblock \enquote{{Statistical Design and Analysis for a ``Biological Effects"
  Study}.}
\newblock {\em Marine Ecology Progress Series\/}, 46, 1, 213--226.

\bibitem[\protect\citename{Cohen, }1963]{cohen:1963}
Cohen, A. (1963).
\newblock \enquote{{Estimation in Mixtures of Discrete Distributions.}}
\newblock In {\em Proceedings of the International Symposium on Discrete
  Distributions, Montreal\/},  373--378.

\bibitem[\protect\citename{Crainiceanu et~al., }2005]{crainiceanu2005bayesian}
Crainiceanu, C., Ruppert, D., and Wand, M. (2005).
\newblock \enquote{{Bayesian Analysis for Penalized Spline Regression Using
  WinBUGS}.}
\newblock {\em Journal of Statistical Software\/}, 14, 14, 1--24.

\bibitem[\protect\citename{Cressie and Wikle, }2011]{cressiewikle2011}
Cressie, N. and Wikle, C. (2011).
\newblock {\em {Statistics for Spatio-Temporal Data}\/}.
\newblock John Wiley and Sons.

\bibitem[\protect\citename{Dagne, }2010]{dagne2010bayesian}
Dagne, G. (2010).
\newblock \enquote{{Bayesian Semiparametric Zero-Inflated Poisson Model for
  Longitudinal Count Data}.}
\newblock {\em Mathematical Biosciences\/}, 224, 2, 126--130.

\bibitem[\protect\citename{Fahrmeir and Echavarr\'{i}a,
  }2006]{fahrmeir2006structured}
Fahrmeir, L. and Echavarr\'{i}a, L. (2006).
\newblock \enquote{{Structured Additive Regression for Overdispersed and
  Zero-Inflated Count Data}.}
\newblock {\em Applied Stochastic Models in Business and Industry\/}, 22, 4,
  351--369.

\bibitem[\protect\citename{Fahrmeir and Tutz, }2001]{fahrmeir2001multivariate}
Fahrmeir, L. and Tutz, G. (2001).
\newblock {\em {Multivariate Statistical Modelling Based on Generalized Linear
  Models}\/}.
\newblock Springer Verlag.

\bibitem[\protect\citename{Gimenez et~al., }2006]{gimenez2006semiparametric}
Gimenez, O., Crainiceanu, C., Barbraud, C., Jenouvrier, S., and Morgan, B.
  (2006).
\newblock \enquote{{Semiparametric Regression in Capture-Recapture Modeling}.}
\newblock {\em Biometrics\/}, 62, 3, 691--698.

\bibitem[\protect\citename{Hall, }2000]{hall2000zero}
Hall, D. (2000).
\newblock \enquote{{Zero-Inflated Poisson and Binomial Regression with Random
  Effects: A Case Study}.}
\newblock {\em Biometrics\/}, 56, 4, 1030--1039.

\bibitem[\protect\citename{Holan et~al., }2008]{holan2008semiparametric}
Holan, S., Wang, S., Arab, A., Sadler, E., and Stone, K. (2008).
\newblock \enquote{{Semiparametric Geographically Weighted Response Curves with
  Application to Site-Specific Agriculture}.}
\newblock {\em {Journal of Agricultural, Biological, and Environmental
  Statistics}\/}, 13, 4, 424--439.

\bibitem[\protect\citename{Johnson et~al., }1997]{johnson1997discrete}
Johnson, N., Kotz, S., and Balakrishnan, N. (1997).
\newblock {\em {Discrete Multivariate Distributions}\/}, vol. 157.
\newblock Wiley New York.

\bibitem[\protect\citename{Kammann and Wand, }2003]{kammann2003geoadditive}
Kammann, E. and Wand, M. (2003).
\newblock \enquote{{Geoadditive Models}.}
\newblock {\em Journal of the Royal Statistical Society: Series C (Applied
  Statistics)\/}, 52, 1, 1--18.

\bibitem[\protect\citename{Kocherlakota and Kocherlakota,
  }1992]{kocherlakota1992bivariate}
Kocherlakota, S. and Kocherlakota, K. (1992).
\newblock {\em {Bivariate Discrete Distributions}\/}.
\newblock Chapman \& Hall/CRC.

\bibitem[\protect\citename{Lam et~al., }2006]{lam2006semiparametric}
Lam, K., Xue, H., and Bun~Cheung, Y. (2006).
\newblock \enquote{{Semiparametric Analysis of Zero-Inflated Count Data}.}
\newblock {\em Biometrics\/}, 62, 4, 996--1003.

\bibitem[\protect\citename{Lambert, }1992]{lambert1992zero}
Lambert, D. (1992).
\newblock \enquote{{Zero-Inflated Poisson Regression, with an Application to
  Defects in Manufacturing}.}
\newblock {\em Technometrics\/}, 34, 1, 1--14.

\bibitem[\protect\citename{Li et~al., }1999]{li1999multivariate}
Li, C., Lu, J., Park, J., Kim, K., Brinkley, P., and Peterson, J. (1999).
\newblock \enquote{{Multivariate Zero-Inflated Poisson Models and Their
  Applications}.}
\newblock {\em Technometrics\/}, 41, 1, 29--38.

\bibitem[\protect\citename{Liu et~al., }2010]{liunonparametric}
Liu, H., Ciannelli, L., Decker, M., Ladd, C., and Chan, K. (2010).
\newblock \enquote{{Nonparametric Threshold Model of Zero-Inflated
  Spatio-Temporal Data with Application to Shifts in Jellyfish Distribution}.}
\newblock {\em Journal of Agricultural, Biological and Environmental
  Statistics\/},  1--17.

\bibitem[\protect\citename{Majumdar et~al., }2010]{majumdar2010bivariate}
Majumdar, A., Gries, C., Walker, J., and Grimm, N. (2010).
\newblock \enquote{{Bivariate Zero-Inflated Regression for Count Data: A
  Bayesian Approach with Application to Plant Counts}.}
\newblock {\em The International Journal of Biostatistics\/}, 6, 1.

\bibitem[\protect\citename{Martin et~al., }2005]{martin2005zero}
Martin, T., Wintle, B., Rhodes, J., Kuhnert, P., Field, S., Low-Choy, S., Tyre,
  A., and Possingham, H. (2005).
\newblock \enquote{{Zero Tolerance Ecology: Improving Ecological Inference by
  Modelling the Source of Zero Observations}.}
\newblock {\em Ecology Letters\/}, 8, 11, 1235--1246.

\bibitem[\protect\citename{Minami et~al., }2007]{minami2007modeling}
Minami, M., Lennert-Cody, C., Gao, W., and Rom{\'a}n-Verdesoto, M. (2007).
\newblock \enquote{{Modeling Shark by Catch: The Zero-Inflated Negative
  Binomial Regression Model with Smoothing}.}
\newblock {\em Fisheries Research\/}, 84, 2, 210--221.

\bibitem[\protect\citename{Musio et~al., }2010]{musio2010bayesian}
Musio, M., Sauleau, E., and Buemi, A. (2010).
\newblock \enquote{{Bayesian Semi-Parametric ZIP Models with Space-Time
  Interactions: An Application to Cancer Registry Data}.}
\newblock {\em Mathematical Medicine and Biology\/}, 27, 2, 181.

\bibitem[\protect\citename{Ntzoufras, }2009]{ntzoufras2009}
Ntzoufras, I. (2009).
\newblock {\em {Bayesian Modeling Using WinBUGS}\/}.
\newblock Wiley: Hoboken, New Jersey.

\bibitem[\protect\citename{Robert and Casella, }2004]{robert2004monte}
Robert, C. and Casella, G. (2004).
\newblock {\em {Monte Carlo Statistical Methods}\/}.
\newblock Springer-Verlag.

\bibitem[\protect\citename{Royle and Dorazio, }2008]{royle2008hierarchical}
Royle, J. and Dorazio, R. (2008).
\newblock {\em {Hierarchical Modeling and Inference in Ecology: The Analysis of
  Data from Populations, Metapopulations and Communities}\/}.
\newblock Academic Press.

\bibitem[\protect\citename{Ruppert et~al., }2003]{ruppert2003semiparametric}
Ruppert, D., Wand, M., and Carroll, R. (2003).
\newblock {\em {Semiparametric Regression}\/}.
\newblock Cambridge University Press.

\bibitem[\protect\citename{Ruppert et~al., }2009]{ruppert2009semiparametric}
--- (2009).
\newblock \enquote{{Semiparametric Regression During 2003--2007}.}
\newblock {\em Electronic Journal of Statistics\/}, 3, 1193.

\bibitem[\protect\citename{Schmidt and Rodriguez, }2011]{schmidt2011}
Schmidt, A. and Rodriguez, M. (2011).
\newblock \enquote{{Modelling Multivariate Counts Varying Continuously in
  Space}.}
\newblock In {\em Bayesian Statistics 9\/}, eds. J.~Bernardo, M.~Bayarri,
  J.~Berger, A.~Dawid, D.~Heckerman, S.~{A. F. M.}, and M.~West. Oxford: Oxford
  University Press.

\bibitem[\protect\citename{Skellam, }1952]{skellam1952studies}
Skellam, J. (1952).
\newblock \enquote{{Studies in Statistical Ecology: Spatial Pattern}.}
\newblock {\em Biometrika\/}, 39, 3-4, 346.

\bibitem[\protect\citename{Spiegelhalter et~al.,
  }2002]{spiegelhalter2002bayesian}
Spiegelhalter, D., Best, N., Carlin, B., and Van Der~Linde, A. (2002).
\newblock \enquote{{Bayesian Measures of Model Complexity and Fit}.}
\newblock {\em Journal of the Royal Statistical Society: Series B (Statistical
  Methodology)\/}, 64, 4, 583--639.

\bibitem[\protect\citename{Tsionas, }2001]{tsionas2001bayesian}
Tsionas, E. (2001).
\newblock \enquote{{Bayesian Multivariate Poisson Regression}.}
\newblock {\em Communications in Statistics-Theory and Methods\/}, 30, 2,
  243--255.

\bibitem[\protect\citename{Ver~Hoef and Jansen, }2007]{ver2007space}
Ver~Hoef, J. and Jansen, J. (2007).
\newblock \enquote{{Space -- Time Zero-Inflated Count Models of Harbor Seals}.}
\newblock {\em Environmetrics\/}, 18, 7, 697--712.

\bibitem[\protect\citename{Wang et~al., }2002]{wang2002zero}
Wang, K., Yau, K., and Lee, A. (2002).
\newblock \enquote{{A Zero-Inflated Poisson Mixed Model to Analyze Diagnosis
  Related Groups with Majority of Same-Day Hospital Stays}.}
\newblock {\em Computer Methods and Programs in Biomedicine\/}, 68, 3,
  195--203.

\bibitem[\protect\citename{Welsh et~al., }1996]{welsh1996modelling}
Welsh, A., Cunningham, R., Donnelly, C., and Lindenmayer, D. (1996).
\newblock \enquote{{Modelling the Abundance of Rare Species: Statistical Models
  for Counts with Extra Zeros}.}
\newblock {\em Ecological Modelling\/}, 88, 1-3, 297--308.

\bibitem[\protect\citename{Wikle, }2003]{wikle2003hierarchicala}
Wikle, C. (2003).
\newblock \enquote{{Hierarchical Models in Environmental Science}.}
\newblock {\em International Statistical Review\/}, 71, 2, 181--199.

\bibitem[\protect\citename{Wikle and Anderson, }2003]{wikle2003ZIP}
Wikle, C. and Anderson, C. (2003).
\newblock \enquote{{Climatological Analysis of Tornado Report Counts Using a
  Hierarchical Bayesian Spatio-temporal Model}.}
\newblock {\em Journal of Geophysical Research-Atmospheres\/}, 108, D24, 9005,
  doi:10.1029/2002JD002806.

\bibitem[\protect\citename{Wildhaber et~al., }2011]{wildhaber2011}
Wildhaber, M.~L., Gladish, D., and Arab, A. (2011).
\newblock \enquote{{Distribution And Habitat Use Of The Missouri River And
  Lower Yellowstone River Benthic Fishes From 1996 To 1998: A Baseline For Fish
  Community Recovery}.}
\newblock {\em River Research and Applications\/}, In Press.

\bibitem[\protect\citename{Zhao et~al., }2006]{zhao2006general}
Zhao, Y., Staudenmayer, J., Coull, B., and Wand, M. (2006).
\newblock \enquote{{General Design Bayesian Generalized Linear Mixed Models}.}
\newblock {\em Statistical Science\/}, 21, 1, 35--51.

\end{thebibliography}
%
%
%\baselineskip=18pt
%
%
%
%
%
%
%
%
%
%
%
%
%
%
%\clearpage
%\pagebreak\newpage \thispagestyle{empty}

\begin{footnotesize}
\begin{center}
\begin{threeparttable}
\begin{tabular}{|l|l|l|l|l|}
\hline
\multicolumn{2}{|c|}{Model} & \multicolumn{2}{c|}{Description} & \multicolumn{1}{c|}{DIC}  \\
\cline{3-4}
\multicolumn{1}{|l}{} &  & \multicolumn{1}{c|}{Linear} & \multicolumn{1}{c|}{Nonlinear} &   \\
\hline\hline
M1 & $\tilde{\lambda}_1$ & segment; macrohab; substrate; gear; year  & temp;  depth; lturb; conduct & 385, 260   \\
 %&  & & lturbidity; conductivity  &    \\
\cline{2-4}
 & $\tilde{\lambda}_2$ & segment; macrohab; substrate; gear; year   & temp;  depth; lturb; conduct &    \\
 %&  &  & lturbidity;  conductivity   &    \\
\cline{2-4}
 & $\tilde{\lambda}_3$ & segment; macrohab; substrate; gear; year  & temp;  depth; lturb; conduct&    \\
 %&  &   &  lturbidity; conductivity  &    \\
\hline\hline
M2 & $\tilde{\lambda}_1$ & segment; macrohab; substrate; gear; year; conduct; temp  & lturb; depth; & 385, 440   \\
 %&  & temp; conduct &  &    \\
\cline{2-4}
 & $\tilde{\lambda}_2$ & segment; macrohab; substrate; gear; year; conduct; temp & lturb; depth; &    \\
% &  & water temperature; conductivity; &  &    \\
\cline{2-4}
 & $\tilde{\lambda}_3$ & segment; macrohab; substrate; gear; year; conduct; temp & \multicolumn{1}{c|}{-} &    \\
&  &   lturb; depth &   &    \\
\hline\hline
M3 & $\tilde{\lambda}_1$ & segment; macrohab; substrate; gear; year; & lturb; depth & 384, 850   \\
\cline{2-4}
 & $\tilde{\lambda}_2$ & segment; macrohab; substrate; gear; year; & lturb; depth &    \\
\cline{2-4}
 & $\tilde{\lambda}_3$ & segment; macrohab; substrate; gear; year; lturb; depth& \multicolumn{1}{c|}{-} &    \\
\hline\hline
M4 & $\tilde{\lambda}_1$ & segment; macrohab; gear & lturb; depth & 386, 670   \\
\cline{2-4}
 & $\tilde{\lambda}_2$ & segment; macrohab; gear & lturb; depth; &    \\
\cline{2-4}
 & $\tilde{\lambda}_3$ & segment; macrohab; gear; year & \multicolumn{1}{c|}{-} &    \\
\hline\hline
M5 & $\tilde{\lambda}_1$ & segment; macrohab; substrate; gear; year & temp; depth; lturb; conduct& 385, 110   \\
 %&  &   &   lturbidity; conductivity  &    \\
\cline{2-4}
 & $\tilde{\lambda}_2$ & segment; macrohab; substrate; gear; year & temp; depth; lturb; conduct &    \\
%  &  &   &  lturbidity; conductivity  &    \\
\cline{2-4}
 & $\tilde{\lambda}_3$ & segment; macrohab; substrate; gear; year; temp; depth; & \multicolumn{1}{c|}{-} &    \\
&  &  lturb; conduct &    &    \\
\hline\hline
M6 & $\tilde{\lambda}_1$ & segment; macrohab; substrate; gear; year; temp; depth; & \multicolumn{1}{c|}{-} & 387, 230   \\
&  &  lturb; conduct &    &    \\
\cline{2-4}
 & $\tilde{\lambda}_2$ & segment; macrohab; substrate; gear; year; temp; depth; & \multicolumn{1}{c|}{-} &    \\
&  &  lturb; conduct &    &    \\
\cline{2-4}
 & $\tilde{\lambda}_3$ & segment; macrohab; substrate; gear; year; & temp; depth; lturb; conduct&    \\
%  &  &   &  lturbidity; conductivity  &    \\
\hline\hline
M7 & $\tilde{\lambda}_1$ & segment; macrohab; substrate; gear; year; temp; depth; & \multicolumn{1}{c|}{-} & 387, 630   \\
&  & lturb; conduct &  &    \\
\cline{2-4}
 & $\tilde{\lambda}_2$ & segment; macrohab; substrate; gear; year; temp; depth; & \multicolumn{1}{c|}{-} &    \\
&  & lturb; conduct &  &    \\
\cline{2-4}
 & $\tilde{\lambda}_3$ & segment; macrohab; substrate; gear; year; temp; depth; & \multicolumn{1}{c|}{-} &    \\
 &  & lturb; conduct &  &    \\
\hline\hline
\end{tabular}
 \caption{For each log-intensity model, linear and nonlinear covariates are identified. The following notation was used in the table:  lturb= log(turbidity), temp=water temperature, conduct=conductivity, and macrohab=macrohabitats.} \label{tab:DIC}
\end{threeparttable}
\end{center}
\end{footnotesize}

\clearpage\pagebreak\newpage \thispagestyle{empty}

\begin{table}
\begin{footnotesize}
\begin{center}
\begin{tabular}[c]{|l||c|c|c|} \hline
{\bf Param.}&
{\bf channel catfish}&{\bf common carp} & {\bf common process}\\
&{\bf Posterior Mean (95\% CI)}&{\bf Posterior Mean (95\% CI)} & {\bf Posterior Mean (95\% CI)}
 \\ \hline\hline
        Seg3        &        -2.8977 (-4.0473,  -1.6142)&    0.5730  (-0.1712,  1.3153)&   -0.6527  (-1.6418,  0.3045)\\\hline
        Seg5        &        -2.5796 (-3.4236,  -1.7467)&    0.1924  (-0.4510,  0.8286)&   -0.7982  (-1.6160,  0.0006)\\\hline
        Seg9        &        -1.6931 (-2.3611,  -1.0417)&    0.2243  (-0.4124,  0.8627)&   -0.7336  (-1.6109,  0.1172)\\\hline
        Seg15       &        -1.7297 (-2.4873,  -0.9597)&    0.2340  (-0.4479,  0.9211)&   -0.7650  (-1.6191,  0.0357)\\\hline
        Seg17       &         0.0624 (-0.7645,   0.8735)&   -1.0825  (-1.7457, -0.4198)&   -1.8191  (-2.7630, -0.9727)\\\hline
        Seg19       &        -0.4098 (-1.2070,   0.3949)&   -0.8706  (-1.4945, -0.2428)&   -1.0380  (-1.7323, -0.3670)\\\hline
        Seg22       &         0.4423 (-0.1653,   1.0605)&   -0.8797  (-1.4709, -0.2888)&   -0.4698  (-1.1191,  0.1501) \\\hline
        Seg23       &         0.1924 (-0.3921,   0.7714)&   -0.4901  (-1.0620,  0.0625)&   -0.2710  (-0.8857,  0.3305)\\\hline
        Seg25       &        -0.5881 (-1.1454,  -0.0224)&    0.1891  (-0.3409,  0.7235)&   -0.7966  (-1.4429, -0.1762)\\\hline
        Year96      &        -0.4027 (-0.8393,   0.0365)&   -0.2338  (-0.6302,  0.1581)&   -1.0172  (-1.5447, -0.4924) \\\hline
        Year97      &         0.0753 (-0.3112,   0.4627)&   -0.1358  (-0.4675,  0.1908)&   -0.3313  (-0.7384,  0.0576)  \\\hline
        BT          &         1.1022 (0.4357,    1.7808)&   -4.0910  (-5.3555, -2.4229)&   -3.9054  (-5.1592, -2.8494) \\\hline
        BS          &         4.4361 (3.7535,    5.1360)&    0.2288  (-0.4262,  0.8774)&   -2.6593  (-3.4105, -1.9717) \\\hline
        DTN         &        -2.5867 (-3.4899,  -1.6776)&   -4.6965  (-5.6991, -3.7468)&   -4.0462  (-5.2319, -2.9813) \\\hline
        BEND        &         0.8567 (0.2506,    1.4599)&   -0.7176  (-1.1309, -0.2940)&    0.7602  (0.2961,   1.2191) \\ \hline
        SCC         &         0.3071 (-0.3347,   0.9511)&   -0.9172  (-1.4212, -0.4202)&    0.1124  (-0.4598,  0.6776) \\\hline
        SCN         &        -0.0788 (-0.9383,   0.7769)&    0.0417  (-0.5926,  0.6756)&   -0.1039  (-1.0000,  0.7414)  \\\hline
        Gravel      &        -0.1088 (-0.8505,   0.6157)&   -0.1888  (-0.8079,  0.4266)&   -0.2771  (-1.1363,  0.5704)  \\\hline
        Sand        &        -0.0093 (-0.4739,   0.4482)&   -0.3235  (-0.7265,  0.0806)&   -0.4431  (-0.9843,  0.0815)  \\\hline
\hline\hline
\end{tabular}
\end{center}
\end{footnotesize}
\caption{ \baselineskip=10pt Posterior mean and 95\% Credible Intervals (CI's) for the log-intensity model parameters for channel catfish, common carp, and the common process.}
\label{tab:loglin}
\end{table}

\clearpage\pagebreak\newpage \thispagestyle{empty}

\begin{table}
\begin{footnotesize}
\begin{center}
\begin{tabular}[c]{|l||c|c|c|}
\hline
{\bf Param.}& {\bf $p_1$}  & {\bf $p_2$}   & {\bf $p_3$} \\
&{\bf Posterior Mean (95\% CI)} & {\bf Posterior Mean (95\% CI)}  & {\bf Posterior Mean (95\% CI)}
 \\ \hline\hline 
        intercept   &        -4.7028  (-9.0578,     0.0402)&       1.4526  (-2.5284,     5.1261)&       6.8226  (3.5612,  10.5133)    \\ \hline
        Seg3        &        -0.8611  (-6.6299,     3.2029)&      -0.0640  (-6.6842,     5.3230)&       2.3359  (-0.8227,  5.4368)    \\ \hline
        Seg5        &         2.0933  (-1.5951,     4.9540)&       2.3959  (-2.1738,     6.1621)&       1.3283  (-1.9495,  4.7725)    \\ \hline
        Seg9        &         3.3999  (-2.3802,     6.9207)&      -0.9648  (-7.2835,     4.2919)&       5.2317  (1.9344,   9.1258)    \\ \hline
        Seg15       &         5.8175  ( 2.9501,     8.9167)&      -0.5501  (-6.5833,     4.3153)&       1.4903  (-2.3358,  4.9361)    \\ \hline
        Seg17       &         0.5610  (-2.8393,     3.5829)&       2.2280  (-0.9609,     5.8953)&      -2.5977  (-5.7459,  0.5701)    \\ \hline
        Seg19       &        -2.4769  (-7.4011,     1.3635)&       0.6779  (-2.7600,     4.1631)&      -1.8595  (-4.6993,  0.8400)    \\ \hline
        Seg22       &         0.7295  (-3.3883,     3.5429)&      -0.5137  (-5.5076,     3.1892)&       1.7871  (-1.2141,  4.9684)     \\ \hline
        Seg23       &         1.5955  (-0.8711,     3.9552)&       0.2700  (-4.3055,     3.6490)&       0.2901  (-2.2090,  3.0414)    \\ \hline
        Seg25       &        -2.0397  (-6.4379,     1.4508)&      -1.4959  (-6.6522,     2.3610)&       0.4252  (-2.0641,  2.8578)    \\ \hline
        Year96      &         0.1058  (-2.7153,     2.3998)&       0.4619  (-2.9069,     3.4959)&      -0.8980  (-2.6601,  0.9159)    \\ \hline
        Year97      &         0.3047  (-1.4503,     2.2170)&       1.7342  (-0.7826,     4.9601)&      -1.1757  (-2.8134,  0.4413)   \\ \hline
        BT          &         3.1982  (-6.5612,     9.9808)&      -1.4747  (-7.1871,     3.4423)&       0.9154  (-2.7345,  5.5469)    \\ \hline
        BS          &         0.1692  (-6.5510,     6.4718)&      -0.9743  (-6.6982,     5.4512)&       2.8631  (-1.2833,  8.0389)    \\ \hline
        DTN         &         1.2121  (-2.7454,     5.7970)&      -1.5209  (-5.5394,     3.0484)&      -8.7597  (-11.7531, -6.3251)    \\ \hline
        BEND        &         0.2153  (-3.1624,     3.8066)&      -2.9688  (-6.0879,    -0.1890)&      -1.0458  (-4.2099,   1.6557)    \\ \hline
        SCC         &        -0.0985  (-4.1622,     4.2813)&      -3.7038  (-7.9150,     0.3265)&       0.4479  (-2.8572,   3.7453)    \\ \hline
        SCN         &         0.2414  (-6.2622,     6.3344)&      -1.1618  (-6.3401,     3.8653)&       1.5549  (-3.8720,   7.3595)    \\ \hline
        Gravel      &         0.3548  (-3.4301,     4.1525)&      -1.2147  (-6.1241,     2.9013)&       0.9386  (-2.0465,   4.2275)    \\ \hline
        Sand        &         0.4618  (-2.7649,     4.0411)&      -2.2307  (-7.4101,     1.3348)&      -0.0256  (-2.7653,   2.4096)    \\ \hline
\hline\hline
\end{tabular}
\end{center}
\end{footnotesize}
\caption{ \baselineskip=10pt Posterior mean and 95\% Credible Intervals (CI's) for the multinomial logit regression model parameters for $p_1$, $p_2$ and $p_3$ with respect to the baseline probability $p_0$.}
\label{tab:logp}
\end{table}

\clearpage\pagebreak\newpage \thispagestyle{empty}

\clearpage\pagebreak\newpage \thispagestyle{empty}

\end{document}